\documentclass[12pt,a4paper]{article}
\usepackage[margin=1in]{geometry}
\usepackage[T1]{fontenc}
\usepackage[scaled]{helvet}

\usepackage{amsmath, amssymb, amsthm}
\usepackage{graphicx}
\usepackage{xcolor}
\usepackage{listings}
\usepackage{caption}
\usepackage[super,sort&compress]{natbib}

\usepackage[hidelinks]{hyperref}
\usepackage{placeins}
\usepackage{float}
\usepackage{pdflscape}
\usepackage{booktabs}
\usepackage{algorithm}
\usepackage{algpseudocode}
\usepackage{subcaption}
\usepackage{multirow}
\usepackage{tikz}
\usetikzlibrary{arrows.meta, positioning}

\newcommand\independent{\protect\mathpalette{\protect\independenT}{\perp}}
\def\independenT#1#2{\mathrel{\rlap{$#1#2$}\mkern2mu{#1#2}}}

\hypersetup{pdfauthor={Yusufhan Balci, Mark A. van de Wiel, Jeroen Hoogland},pdftitle={B-SLIM for heterogeneous treatment effects}}
\begin{document}

\title{B-SLIM: BART Sparse Linear Interactions Model for Heterogeneous Treatment Effects}
\author{\normalsize Yusufhan Balci$^{1,2}$, Mark A. van de Wiel$^{1,2}$ and Jeroen Hoogland$^{1,2}$}
\date{}
\maketitle
\begin{center}
\small
$^1$Amsterdam UMC, Vrije Universiteit Amsterdam, Department of Epidemiology and Data Science, Amsterdam, The Netherlands\\[3pt]
$^2$Amsterdam Public Health research institute, Methodology, Amsterdam, The Netherlands
\end{center}
\noindent\textbf{Correspondence:} Yusufhan Balci, Department of Epidemiology and Data Science, Amsterdam UMC, Amsterdam, The Netherlands. Email: \href{mailto:H.y.h.balci@amsterdamumc.nl}{H.y.h.balci@amsterdamumc.nl}.

\smallskip\noindent\textbf{Short title:} B-SLIM for heterogeneous treatment effects


\begin{abstract}
Estimating covariate-dependent treatment effects is a central challenge in personalized medicine. Existing approaches often emphasize either fully parametric models, risking misspecification, or flexible machine learning methods, which may require large samples to detect the small treatment effect heterogeneity common in medical settings. We propose B-SLIM (BART Sparse Linear Interaction Model), a semi-parametric model for causal inference that combines a flexible Bayesian Additive Regression Trees (BART) prognostic component with a sparse linear treatment effect component. This hybrid structure combines flexible prognosis modelling with interpretable treatment effects. The treatment effect component includes main effects and pairwise interactions, regularized through a global--local shrinkage prior. The model allows covariates to contribute to both components and provides direct inference on treatment effect modifiers, conditional mean potential outcomes, and conditional average treatment effects (CATEs). In simulations, B-SLIM outperforms a fully non-parametric alternative when main effects and pairwise interactions adequately represent the treatment effect function, whereas the non-parametric model performs better under strong non-linearity. On the nonlinear Infant Health and Development Program (IHDP) benchmark, the two Bayesian methods are competitive in predictive accuracy and ranking, and both achieve lower mean CATE error than the DR-learner, S-learner, and R-learner implementations considered. In the CANVAS trials, both methods yield similar rankings, while B-SLIM identifies several important drivers of treatment effect heterogeneity. B-SLIM enables interpretable inference on treatment effect parameters and patient-specific prediction of conditional mean potential outcomes, while remaining competitive at sample sizes relevant to medical studies.
\end{abstract}


\clearpage
\section{Introduction}
A treatment that is beneficial on average for a population might have varying degrees of benefit for different subgroups; a drug could benefit the majority of patients while causing severe side-effects for others. In hormone-receptor-positive, HER2-negative, node-negative breast cancer, genomic scores such as Oncotype DX are used to decide which patients are likely to benefit from chemotherapy and which can safely forgo it.\citep{sparano2018adjuvant} Assessing this individual treatment benefit matters for drug development, clinical decision-making, and policy recommendations. These recommendations rely on four key questions regarding treatment effect heterogeneity: ``Who benefits from treatment?", ``Why do these individuals (not) benefit?" (i.e., which covariates drive the treatment effect), ``How certain are we about the expected benefit for someone with these characteristics?", and ``What would have happened if this individual was (not) treated?".

The question of ``What if?" forms the basis of the causal inference literature. The potential outcomes framework \citep{rubin1974estimating} conceptualizes individuals as having two potential outcomes in a binary treatment setting: one under treatment and one under control. Because we only observe one outcome under a single treatment regime, the counterfactual remains missing. Assumptions and modelling are therefore required to infer the Average Treatment Effect (ATE) and the Conditional Average Treatment Effect (CATE), which denotes the expected treatment effect conditional on specific individual characteristics. In applied clinical research, subgroup analyses are routinely conducted to investigate whether certain covariates significantly modify the treatment effect. However, these repeated analyses often suffer from a lack of statistical power and prevalent Type-I errors.\citep{kent2020predictive}

To overcome this lack of power and directly model the CATE, Tian et~al.\citep{tian2014simple} modelled the contrast between control and treated groups using a regularized linear model with interactions, treating the prognostic function as a nuisance parameter. Because the surface of the CATE is often smoother and easier to estimate than the separate potential outcome surfaces,\citep{curth2021inductive} modern causal machine learning methods focus on directly estimating this treatment contrast. Key subsequent developments include causal trees \citep{athey2016recursive} and Causal Random Forests (CRF),\citep{wager2018estimation} which estimate the CATE using a tree ensemble inspired by the Random Forests algorithm.\citep{breiman2001random} Although asymptotic guarantees have been established for CRF,\citep{wager2018estimation} its reliance on deep decision trees can lead to poor performance in settings with limited sample sizes.\citep{hahn2020bayesian} Across different causal inference competitions, models using Bayesian Additive Regression Trees (BART) \citep{chipman2010bart} priors frequently outperform other methods.\citep{dorie2019automated} The inherent regularization on tree depth and leaf nodes renders BART less susceptible to overfitting while maintaining the flexibility of an ensemble model. Building on this, Hahn et~al.\citep{hahn2020bayesian} propose Bayesian Causal Forests (BCF), which models the outcome space using two BART priors---one for the prognostic function and one for the treatment effect function---and provides posterior distributions for individual CATEs.

Although these Bayesian tree-based methods handle non-linear outcome relationships and provide credible intervals, two difficulties remain for practitioners. First, uncertainty quantification for these CATEs is difficult, yet it is needed if the estimates are to inform policy. These models provide a posterior for the CATE, but the frequentist coverage of its credible intervals can depend on regularization and the treatment effect structure. Second, black-box tree ensembles provide limited insight into \emph{which} specific covariates are driving the treatment effect heterogeneity. Although post-hoc explainability techniques, such as fitting an additional surrogate machine learning model to the CATE predictions, can be employed,\citep{svensson2026overview} this introduces the overhead of secondary modelling and additional estimation variance.

To address these points and answer the four questions above, we propose B-SLIM (BART Sparse Linear Interaction Model) for causal inference. The prognostic baseline risk is modelled non-parametrically using BART, while the heterogeneity of the treatment effect is modelled parametrically using a linear formulation with main effects and interactions. The semi-parametric BART model of Zeldow et~al.\citep{zeldow2019semiparametric} combines the same two ingredients: a BART nuisance component and a parametric component for the treatment and its effect modifiers. Their structural mean model also allows an effect modifier to contribute to prognosis through a linear term. Related work by Prado et~al.\citep{prado2025shared} allows covariates to be shared between the linear and BART components, modifying the tree updates to address identifiability. As they use non-sparse priors on the regression coefficients one may be forced to reduce the parameter space by selecting potential effect modifiers beforehand and ignoring two-way interactions between those. In practice effect modifiers are rarely known beforehand. We therefore allow the same covariates to enter the prognostic and treatment functions, and consider the full covariate space, including pairwise interactions, in the treatment effect function and regularize the resulting coefficient space. This space becomes large quickly, which is why we employ a global--local shrinkage prior.\citep{carvalho2010horseshoe} Such a prior combines a global scale, which pulls all coefficients towards zero, with a local scale for each coefficient, which lets a strong signal escape this shrinkage. The horseshoe prior is the best known example. It shrinks noise terms towards zero while allowing relevant signals to escape strong shrinkage, thereby limiting shrinkage bias.\citep{van2014horseshoe} The sampling of our model is a modification of the Bayesian backfitting algorithm used by Hahn et~al.\citep{hahn2020bayesian} and Hastie and Tibshirani.\citep{hastie2000bayesian} Finally, we demonstrate that our framework allows straightforward computation of Shapley values, which quantify the importance of each covariate for predicting the CATE, including the contribution of interaction terms. To summarize, our contribution is the combination of a regularized BART prognostic component with a parametric treatment component, regularized by global--local shrinkage, that allows appropriate inference and interpretation.

We evaluate the model in simulations, the semi-synthetic Infant Health and Development Program (IHDP) benchmark, and an application to the CANVAS Program. The simulations examine settings in which the treatment-effect structure is aligned with the model and settings in which it is misspecified. IHDP combines real covariate data with 100 simulated potential-outcome scenarios, so individual treatment effects are known. The CANVAS application examines canagliflozin in patients with type 2 diabetes at high cardiovascular risk. Section 2 presents the model, Section 3 describes the study settings, Section 4 reports the simulation and IHDP results, and Section 5 presents the CANVAS application before the discussion and conclusion.

\section{Methodology}
\label{sec:method}
B-SLIM combines the flexible prognostic estimation of BART with the interpretable inference of a sparse linear interaction model for treatment effect heterogeneity. The BART prior allows the prognostic component to capture non-linearities and interactions. The sparse global--local shrinkage prior on the linear treatment effect component allows parameter inference and interpretation for the drivers of the treatment effect.

\subsection{Causal Framework}
Our main interest is in predicting and making inference on the conditional mean potential outcomes $\operatorname{E}\{y_i(0)\mid x_i\}$ and $\operatorname{E}\{y_i(1)\mid x_i\}$, and their difference, the CATE $\tau(x_i)$. These describe the expected outcomes under treatment and control for individuals with characteristics $x_i$, rather than identifying both realized outcomes for a single individual. We denote the vector of covariates $x_i$ and the binary treatment assignment $z_i$ for individual $i=1,\ldots,N$. As we only observe one outcome for each individual, we need assumptions to identify these conditional means.\citep{imbens2015causal} These are:
\begin{itemize}
    \item Stable unit treatment value assumption (SUTVA). SUTVA requires that there is no interference between units and that there are no multiple versions of a treatment. This means that the outcome can be written as: $y_i = z_iy_i(1) + (1-z_i)y_i(0)$.
    \item Conditional exchangeability, i.e. no unmeasured confounding given $x_i$;\newline $\{y_i(1),y_i(0)\} \independent z_i \mid x_i$.
    \item Overlap, i.e. $ 0 < P(z_i = 1 | x_i) < 1$. This entails that there always should be some randomness in assignment; in the case that this assumption fails, we would not have enough support to infer the counterfactual.
\end{itemize}

\subsection{B-SLIM}
We model the observed outcome $y_i$ for individual $i=1,\dots,N$ as the sum of a prognostic function $\mu$ and a treatment effect function $\tau$:
\begin{equation}
\label{main_eq}
    y_{i} = \mu(x_{i}, \hat{\pi}(x_i)) + u_{i}\tau(x_{i}) + \varepsilon_{i}, \quad \varepsilon_i \sim \mathcal{N}(0, \sigma^{2}),
\end{equation}
where $x_i$ is a vector of $p$ covariates, $z_i$ denotes the status of treatment, and $\hat{\pi}(x_i)$ the fitted propensity score. We adopt a standard BART prior for the prognostic function $\mu(\cdot)$, allowing it to capture non-linearities and interactions in the baseline risk. To address regularization-induced confounding (RIC), where strongly regularized priors might not sufficiently adjust for confounding due to shrinkage-induced bias, we include the fitted propensity score $\hat{\pi}(x_i)$ as a covariate in $\mu$, linking priors of outcome and assignment mechanisms.\citep{hahn2020bayesian, linero2024nonparametric} In observational data, we estimate the propensity score with a Bayesian logistic regression with main effects and pairwise interactions. In a randomized trial, the assignment probability is known from the design and need not be estimated; a constant assignment probability adds no information as a splitting variable. To facilitate direct inference on effect modifiers, we specify the CATE, $\tau(x_i)$, as a linear model with interactions:
\begin{equation}
\label{linear_tau}
    \tau(x_{i}) = \alpha + \sum_{j=1}^{p} \beta_{j}\, x_{i,j} + \sum_{j<k} \beta_{j,k}\, x_{i,j}\,x_{i,k},
\end{equation}
where $\beta_j$ and $\beta_{j,k}$ respectively represent the main and interaction effects. Here $p$ counts the covariate columns after encoding. Products between two categorical contrast columns are excluded, with their coefficients set to zero.

Let $u_i$ denote the fixed treatment coding, with $u_i=z_i$ for $0/1$ coding and $u_i=z_i-1/2$ for symmetric coding. The binary encoding of $z_i$ (e.g., $0/1$ versus $-0.5/0.5$) implies an assumption about how the prior variability is distributed between the treatment and control mean outcome functions, and Hahn et~al.\citep{hahn2020bayesian} propose to estimate the treatment coding. A symmetric treatment coding gives the two mean outcome functions the same prior variability, while $0/1$ coding allows additional prior variability in the treatment arm. We recommend that users choose the appropriate coding based on the problem context. Symmetric coding is preferred when we believe that the variability in both mean outcome functions will be similar; a $0/1$ coding is more appropriate if the active treatment is expected to introduce additional heterogeneity in patient outcomes.

It is also possible to residualise the treatment indicator, using $u_i=z_i-\hat{\pi}(x_i)$ in the linear component in the spirit of Robinson-style orthogonalisation.\citep{robinson1988root} The separate residualised implementation is described in the Supplementary Material (Section \ref{subsec:ortho_supp}); its comparison with the standard model is reported with the simulation results.

\subsubsection{Shrinkage Prior for the Treatment Effect}
\label{subsec:inference}
The number of coefficients in \eqref{linear_tau} grows quadratically with the number of covariates. In order to efficiently shrink nuisance terms towards zero while keeping the bias on the main contributors to the treatment effect minimal, we employ a global--local shrinkage prior \citep{carvalho2010horseshoe} for the coefficients of the main effects and interaction terms. This allows us to search over a large coefficient space while keeping $\tau(x)$ interpretable. The prior hierarchy is defined as:
\begin{align}
    \alpha &\sim \mathcal{N}(0, 10^2), \nonumber\\
    \lambda_{j},\ \lambda_{j,k} &\sim \mathcal{C}^+(0, 1), \nonumber\\
    \lambda_{\mathrm{glob}} &\sim \mathcal{C}^+(0, 1), \nonumber\\
    \beta_{j} \mid \lambda_{j}, \sigma &\sim
      \mathcal{N}\!\left(0,\ \sigma^2 \lambda_{j}^2\right),
      \label{eq:hshc_main}\\
    \beta_{j,k} \mid \lambda_{j,k}, \lambda_{\mathrm{glob}}, \sigma &\sim
      \mathcal{N}\!\left(0,\ \sigma^2 \lambda_{j,k}^2 \lambda_{\mathrm{glob}}^2\right),
      \qquad \text{for included pairs }j<k, \label{eq:hshc_int}\\
    \sigma^2 &\sim \mathrm{InvGamma}(1,\ 0.001). \nonumber
\end{align}
The global shrinkage parameter $\lambda_{\mathrm{glob}}$ sets the general sparsity of the interaction block. The local parameters $\lambda_{j}$ and $\lambda_{j,k}$ allow for the identification of signal independently. We keep the main effects outside this global scale, so that each $\beta_j$ is governed by its own local parameter only, which gives a normal--half-Cauchy scale mixture for $\beta_j/\sigma$, with concentration near zero and heavy tails. The auxiliary variable reparameterization of Makalic and Schmidt\citep{makalic2015simple} permits Gibbs sampling. A coefficient whose $95\%$ credible interval excludes zero provides posterior evidence of effect modification. These are pointwise intervals.

\subsection{Posterior Inference}
We estimate the joint posterior using a Bayesian backfitting algorithm.\citep{hastie2000bayesian} This iterative procedure decouples the estimation of the prognostic and treatment functions by alternating between them using partial residuals. The full conditionals, BART prior, and sampling scheme are given in the Supplementary Material (Section \ref{sec:computational_details} and Algorithm \ref{algorithm backfitting}). At MCMC iteration $m$, the sampling steps are:

\begin{enumerate}
    \item \textbf{Update Prognostic Function:} Compute the partial residuals $R^*_{i} = y_i - u_i \tau^{(m-1)}(x_i)$. Draw $\mu^{(m)}$ from the BART posterior conditional on $R^*$.
    \item \textbf{Update Treatment Function:} Compute the partial residuals $R^{**}_{i} = y_i - \mu^{(m)}(x_i, \hat{\pi}_i)$. Draw the linear parameters for $\tau^{(m)}$ (including $\alpha, \beta, \lambda$) using the appropriate sampler for the shrinkage prior, conditional on $R^{**}$.
    \item \textbf{Update Noise Variance:} Draw $\sigma^{2(m)}$ from its Inverse-Gamma full conditional, using the total residuals and the coefficient-prior terms that depend on $\sigma^2$.
\end{enumerate}

The supplied scripts use four independent chains per model, with 2000 burn-in iterations and 4000 retained draws per chain after 25 grow-from-root initialisation iterations. These settings apply to the simulations, IHDP and CANVAS.

For one simulated data set with $N=500$ (linear treatment effects and non-linear prognosis), the four sequential B-SLIM chains took 2.2 minutes of fitting time on an Intel Core i5-1335U (10 cores, 16 GB RAM; Windows 11, R 4.4.2), excluding data generation and propensity estimation.

The estimation of $\tau(x_i)$ depends on how well the prognostic component accounts for the outcome variation. A better estimate of $\mu(x_i)$ can leave a clearer signal in the partial residuals $R^{**}$, although this does not by itself guarantee consistency or calibrated credible intervals for $\tau(x_i)$.

\subsection{Interpretation}
\label{subsec:shapley}
While coefficient inference identifies statistically significant drivers of heterogeneity, we quantify the total contribution of a covariate to the CATE by marginal Shapley values.\citep{van2024linked} Although often associated with black-box predictors, Shapley values have also been shown to be useful in relatively complex regression settings.\citep{van2024linked} They decompose the predicted $\tau(x_i)$ into contributions of covariates. Unlike for non-parametric models, our linear parametrisation for $\tau(x_i)$ allows analytical computation of these Shapley values \citep{van2024linked}:
\begin{equation}
\label{eq:shapley}
    \phi_{i,j} = \beta_j x^*_{i,j} + \frac{1}{2} \left( \sum_{k:k \neq j} \beta_{j,k} x^*_{ij} x^*_{ik} - \sum_{k:k \neq j} \beta_{j,k} \operatorname{E}[x^*_{i,j}x^*_{i,k}] \right),
\end{equation}
where $x^*$ denotes centered covariates and the coefficients refer to the model expressed in these covariates. This gives a posterior for variable importance that accounts for both the main effect of a covariate and its interactions. From equation \eqref{eq:shapley} it can be written as $\phi_{i,j} = \phi_{i,j,main} + \phi_{i,j,int}$, so that the main-effect and interaction contributions can be separated, with the contributions of contrast columns belonging to the same categorical covariate summed within each individual before taking absolute values.

\subsection{AI Assistance}
\label{subsec:ai_assistance}
OpenAI Codex was used to assist with checking the sampler code, writing scripts to run the simulations on the server, formatting the manuscript in LaTeX, checking references, improving sentence flow and checking for typographical errors. The authors reviewed the analysis code and manuscript and take responsibility for the analyses, interpretation and final manuscript.

\section{Study Settings}
\label{sec:study_settings}
We evaluate the models in two settings where individual treatment effects are known: a simulation study based on Hahn et~al.\citep{hahn2020bayesian} and the semi-synthetic Infant Health and Development Program (IHDP) benchmark. The simulations assess model performance under both correct specification and misspecification of B-SLIM's treatment effect component. The IHDP benchmark provides a complementary assessment using real covariate data and 100 simulated potential-outcome scenarios.

\subsection{Simulation Study}
\label{sec:simulation}
We compare B-SLIM with the fully non-parametric Bayesian Causal Forest (BCF). We focus on these two models because they share a BART prognostic component and both provide predictions for the conditional mean potential outcomes. This comparison examines the trade-off between a regularized parametric and a tree-based treatment function. We estimate the Average Treatment Effect (ATE) and the Conditional Average Treatment Effect (CATE), and examine how both models behave under structural alignment and structural misspecification of the treatment effect.

\subsubsection{Data Generating Process}
\label{subsec:simulation_dgp}
We adopt the data-generating process (DGP) described by Hahn et~al.\citep{hahn2020bayesian} and extend its treatment effect component to include both structurally aligned and misspecified scenarios. We examine model performance across sample sizes $N \in \{250,\allowbreak 500,\allowbreak 750,\allowbreak 1000,\allowbreak 1500,\allowbreak 3000\}$, with 200 independent replicates per cell. The baseline covariates consist of three continuous variables $x_1, x_2, x_3 \sim \mathcal{N}(0,1)$, one binary variable $x_4$ taking values $-1$ and $1$ with equal probability, and a categorical variable $x_5$ uniformly distributed over $\{1,2,3\}$. The five covariates are independent. The prognostic function $\mu(x)$ is defined as:
\[
\mu(x) =
\begin{cases}
    1 + g(x_5) + x_1 x_3, & \text{linear,} \\
    -6 + g(x_5) + 6|x_3 - 1|, & \text{nonlinear,}
\end{cases}
\]
where $g(1) = 2$, $g(2) = -1$, and $g(3) = -4$. Targeted selection into treatment is induced via the propensity score, mimicking medical scenarios where patients with worse prognoses are more likely to be treated:
\[
\pi(x_i) = \max\left(0, \min\left(1, 0.8\Phi\left(\frac{3\mu(x_i)}{s} - 0.5x_1\right) + 0.05 + \frac{v_i}{10}\right)\right),
\]
where $s$ is the sample standard deviation of $\mu(x)$, $\Phi$ is the standard normal CDF, and $v_i \sim \text{Uniform}(0, 1)$. We draw $z_i\sim\operatorname{Bernoulli}\{\pi(x_i)\}$ and generate $y_i=\mu(x_i)+(z_i-0.5)\tau(x_i)+\varepsilon_i$, with independent $\varepsilon_i\sim\mathcal{N}(0,1)$. The categorical variable is represented by two sum-contrast columns. B-SLIM uses symmetric treatment coding, $u_i=z_i-1/2$, matching this data-generating model. BCF uses adaptive treatment coding, as in Hahn et~al.\citep{hahn2020bayesian} For both models, we estimate the propensity score using Bayesian logistic regression with main effects and pairwise interactions and include it in the prognostic component.

\subsubsection{Treatment Effect Scenarios}
\label{subsec:treatment_scenarios}
\begin{figure}
    \centering
    \includegraphics[width=0.65\linewidth]{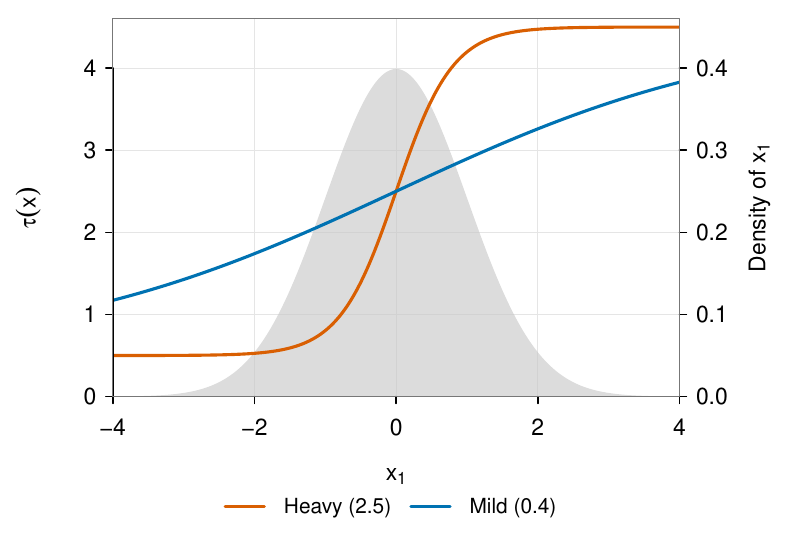}
    \caption{Shape of $\tau(x)$ plotted against $x_1$ for the two saturation scenarios, with coefficients $2.5$ and $0.4$. The shaded area shows the standard normal density of $x_1$ (right axis).}
    \label{fig:saturation-density-plot}
\end{figure}

To examine the trade-off between parametric structure and non-parametric flexibility, we simulate four treatment effect scenarios for the true CATE $\tau(x)$:

\begin{enumerate}
    \item \textbf{Homogeneous Treatment Effect:} The treatment effect is homogeneous for the whole population:
    \[ \tau(x) = 3 \]
    This setting examines whether the models avoid estimating spurious treatment effect heterogeneity.
    \item \textbf{Linear Scenario:} The treatment effect is generated by a simple linear interaction:
    \[ \tau(x) = 1 + 2 x_2 x_4 \]
    This scenario aligns with the structural assumptions of B-SLIM. Alongside the homogeneous scenario, this simulation scenario was presented by Hahn et~al.\citep{hahn2020bayesian} and was also used in other work by Linero.\citep{linero2024nonparametric}
    \item \textbf{Heavy Saturation Scenario:} The treatment effect is driven by a non-linear sigmoid saturation curve based on a continuous biomarker $x_1$:
    \[ \tau(x) = 0.5 + \frac{4}{1 + \exp(-2.5 x_1)} \]
    Patients with low $x_1$ values receive a small baseline benefit, which grows rapidly as $x_1$ increases before capping at a biological saturation limit of 4.5. This scenario represents a functional misspecification for B-SLIM, favouring BCF.
    \item \textbf{Mild Saturation Scenario:} Building on the previous scenario, we model the treatment effect using a sigmoid function with reduced saturation to more closely reflect a linear relationship.  This is defined as:
\[ \tau(x) = 0.5 + \frac{4}{1 + \exp(-0.4 x_1)} \]
In Figure \ref{fig:saturation-density-plot}, $\tau(x)$ is plotted against $x_1$ for the two scenarios, alongside the density of $x_1$ for clarity.
\end{enumerate}

\subsubsection{Performance Measures}
For the ATE and CATE, we report root mean squared error (RMSE), empirical coverage (Cov), and the average length (Len) of the $95\%$ credible intervals. The ATE target in each replicate is the mean of the true CATEs in that sample; ATE RMSE is the square root of the mean squared estimation error across replicates. For the CATE, we first calculate the RMSE across individuals within each replicate and then average over replicates. Coverage and interval length are also averaged over replicates. With 200 independent replicates, the Monte Carlo standard error of an ATE coverage estimate near $95\%$ is about 1.5 percentage points.

We assess how well the models rank individuals by calculating the Spearman correlation between posterior mean CATE estimates and the true CATE. We also evaluate their ability to identify individuals in the top $10\%$, $25\%$, and $50\%$ of the true treatment effect distribution. For each group, an individual is selected only if the lower bound of their $95\%$ credible interval exceeds the corresponding true threshold, which is known in simulation. Precision and sensitivity summarise how well each model identifies these individuals. This criterion reflects both the estimated treatment effects and their uncertainty, whereas Spearman correlation measures ranking alone. Full results for the homogeneous treatment effect scenario and the threshold-based selection analysis are provided in the Supplementary Material.

\subsection{Semi-synthetic IHDP Benchmark}
\label{subsec:ihdp_setting}
The Infant Health and Development Program (IHDP) data are a well-known semi-synthetic benchmark with real covariate data and synthetic potential outcomes under 100 different scenarios.\citep{ihdp1990trial,hill2011bayesian,ihdpbenchmarkdata} The benchmark contains 747 children with 25 covariates, of which five are continuous and one is a four-level categorical variable. Our preprocessing encodes these into 27 columns. Of the children, 139 (18.6\%) are treated. All 100 data sets use the same covariates and the same treatment assignments; only the simulated potential outcomes change. Because the true treatment effect $\tau$ is simulated for each child, CATE estimation accuracy can be assessed against the truth. The benchmark combines an exponential control mean with a linear treated mean: $\mu_0(x)=\exp\{(x+a)^\top b\}$ and $\mu_1(x)=x^\top b-\omega$, so $\tau(x)=\mu_1(x)-\mu_0(x)$. Here $a$ is the covariate offset, $b$ the coefficient vector of the response surface, and $\omega$ is chosen to give an average treatment effect of four among the treated. This is similar in spirit to the saturation settings in Subsection \ref{subsec:treatment_scenarios}, except that the non-linearity is spread over several covariates rather than concentrated in $x_1$. The benchmark and its consequences for comparing CATE estimators are discussed in more detail by Curth et~al.\citep{curth2021really}

The IHDP comparison is restricted to the CATE. That makes it possible to include a DR-learner, an R-learner, and an S-learner alongside B-SLIM and BCF, using ten-fold cross-validation. Within each outer training partition, we obtain cross-fitted propensity estimates from a Bayesian logistic model with main effects and pairwise interactions. These estimates are reused throughout fitting and shared by BCF, B-SLIM, and the DR- and R-learners. In contrast to the synthetic simulations, the IHDP analysis uses out-of-sample evaluation with inner cross-fitting. Outcomes from the held-out data are excluded when fitting the propensity and outcome models. B-SLIM uses symmetric treatment coding to give both mean outcome functions the same prior variability and supplies this propensity estimate to its prognostic component. BCF uses binary treatment coding and adaptive treatment coding.\citep{hahn2020bayesian}

Let $\mu_0(x)$ and $\mu_1(x)$ denote the mean outcomes under control and treatment. The DR-learner estimates these two outcome regressions by cross-fitting a \texttt{SuperLearner} ensemble,\citep{vanderlaan2007super} using the logistic propensity estimates described above. It then builds an augmented inverse-probability-weighted pseudo-outcome \citep{kennedy2020optimal} and fits a second \texttt{SuperLearner} to that pseudo-outcome. The R-learner is a causal forest that uses cross-fitted estimates of $m(x)=\operatorname{E}(Y\mid X=x)$ from the same ensemble, together with the same propensity.\citep{nie2021quasi} The S-learner fits one \texttt{SuperLearner} outcome model to the covariates and treatment indicator and takes the contrast between the treatment arms.\citep{kunzel2019metalearners} The \texttt{SuperLearner} library used for these fits contains a generalized linear model, a penalized generalized linear model, three generalized additive models with two to four degrees of freedom, gradient boosting, and a random forest.

We evaluate the five methods using out-of-sample RMSE for the CATE, Spearman correlation with the true CATE, and the calibration slope. We also report each method's mean rank, by RMSE, within the 100 data sets. Lower RMSE and a lower mean rank are better, a higher Spearman correlation is better, and a calibration slope of one is ideal. The slope is obtained by regressing the true CATE on the estimated CATE with an intercept. For the two Bayesian methods we additionally evaluate the $95\%$ credible intervals for each individual CATE: coverage of the true $\tau_i$, how often the interval excludes zero, and the Type S error among those that do. Because exclusion rates are not comparable when methods have different coverage, we save the complete CATE posterior draws and perform a second, coverage-matched analysis, reported in the Supplementary Material. For each Bayesian model, a scale factor $c$ is chosen using the other IHDP replicates and then evaluated on the held-out replicate. The held-out response surface therefore does not determine its own scale factor, although all replicates share the same children and treatment assignments.

\newpage
\section{Results}
\label{sec:results}

\subsection{Simulation Results}

\subsubsection{Homogeneous Treatment Effect}
In the homogeneous scenario, B-SLIM and BCF both shrink towards homogeneity, with lower CATE RMSE for BCF under linear prognosis (Supplementary Table \ref{tab:hom_scenario}). The true treatment effect is constant, $\tau(x)=3$. In B-SLIM, the shrinkage prior pulls the main-effect and interaction coefficients towards zero, allowing the treatment function to approach a constant. For both models, CATE RMSE decreases with sample size under linear and non-linear prognosis.
\par\smallskip

\subsubsection{Linear Treatment Effect}

In the linear scenario, the true CATE is generated as a simple linear interaction: $\tau(x) = 1 + 2x_2x_4$. Table \ref{tab:default_scenario} shows that B-SLIM generally has lower ATE RMSE and has lower CATE RMSE, shorter interval lengths, and higher CATE coverage than BCF. The absolute difference in CATE performance is most pronounced from $N=250$ to $N=1000$. In this range, BCF often shrinks the treatment effect back towards homogeneity because its tree prior does not represent a linear treatment surface as efficiently. As the sample size increases, BCF estimates the heterogeneity more accurately, although its RMSE for the CATE remains higher. The exact sample size at which this occurs depends on the signal-to-noise ratio. This pattern also holds when the prognostic function $\mu(x)$ is non-linear. At $N=3000$, the CATE credible intervals from B-SLIM are narrower ($0.598$) than those from BCF ($1.299$), while maintaining coverage of $98\%$.

\begin{table}[htbp]
\centering
\caption{Performance Comparison in the \textbf{Default Heterogeneous Scenario} (Linear Interaction CATE). Results are shown for models with Heterogeneous Treatment Effects, comparing Linear and Non-linear prognostic functions ($\mu$).}
\label{tab:default_scenario}
\resizebox{\textwidth}{!}{
\begin{tabular}{lr ccc ccc c ccc ccc}
\toprule
& & \multicolumn{6}{c}{\textbf{BCF}} & & \multicolumn{6}{c}{\textbf{B-SLIM}} \\
\cmidrule{3-8} \cmidrule{10-15}
$\mu(x)$ Form & $N$ & \multicolumn{3}{c}{ATE} & \multicolumn{3}{c}{CATE} & & \multicolumn{3}{c}{ATE} & \multicolumn{3}{c}{CATE} \\
\cmidrule{3-5} \cmidrule{6-8} \cmidrule{10-12} \cmidrule{13-15}
& & RMSE & Cov & Len & RMSE & Cov & Len & & RMSE & Cov & Len & RMSE & Cov & Len \\
\midrule
\multirow{6}{*}{Linear}
& 250 & 0.250 & 0.95 & 0.949 & 0.993 & 0.91 & 3.274 &  & 0.220 & 0.97 & 0.885 & 0.574 & 0.97 & 2.453 \\
& 500 & 0.132 & 0.98 & 0.592 & 0.700 & 0.93 & 2.466 &  & 0.132 & 0.96 & 0.567 & 0.368 & 0.97 & 1.652 \\
& 750 & 0.112 & 0.96 & 0.458 & 0.580 & 0.93 & 2.128 &  & 0.113 & 0.94 & 0.446 & 0.308 & 0.97 & 1.338 \\
& 1000 & 0.098 & 0.96 & 0.388 & 0.520 & 0.94 & 1.909 &  & 0.097 & 0.95 & 0.380 & 0.259 & 0.97 & 1.148 \\
& 1500 & 0.077 & 0.95 & 0.309 & 0.440 & 0.94 & 1.655 &  & 0.077 & 0.94 & 0.305 & 0.213 & 0.97 & 0.932 \\
& 3000 & 0.053 & 0.96 & 0.212 & 0.331 & 0.94 & 1.289 &  & 0.052 & 0.97 & 0.212 & 0.149 & 0.97 & 0.652 \\
\midrule
\multirow{6}{*}{Non-linear}
& 250 & 0.324 & 0.89 & 0.959 & 1.150 & 0.91 & 3.595 &  & 0.235 & 0.94 & 0.829 & 0.601 & 0.96 & 2.432 \\
& 500 & 0.190 & 0.88 & 0.581 & 0.818 & 0.92 & 2.595 &  & 0.158 & 0.91 & 0.535 & 0.372 & 0.97 & 1.601 \\
& 750 & 0.125 & 0.93 & 0.449 & 0.661 & 0.94 & 2.197 &  & 0.108 & 0.96 & 0.423 & 0.278 & 0.98 & 1.270 \\
& 1000 & 0.101 & 0.95 & 0.381 & 0.584 & 0.94 & 1.967 &  & 0.090 & 0.97 & 0.361 & 0.234 & 0.98 & 1.080 \\
& 1500 & 0.085 & 0.91 & 0.299 & 0.491 & 0.93 & 1.665 &  & 0.079 & 0.91 & 0.289 & 0.188 & 0.97 & 0.863 \\
& 3000 & 0.054 & 0.95 & 0.205 & 0.366 & 0.94 & 1.299 &  & 0.049 & 0.97 & 0.201 & 0.129 & 0.98 & 0.598 \\
\bottomrule
\end{tabular}}
\end{table}

\subsubsection{Heavy Saturation}
The comparative advantage for the point estimates shifts in favour of BCF in the heavy saturation scenario (Table \ref{tab:med_sat_scenario}). Here, the treatment effect follows a steep sigmoid curve and is non-linear. The RMSEs are similar at $N=250$. BCF generally has lower CATE RMSE and better coverage because the tree-based treatment function can capture the steep curve by splitting on $x_1$. As $N$ increases, the posterior of B-SLIM concentrates around a biased linear approximation, causing its credible intervals to shrink and coverage to decrease. BCF's advantage under heavy saturation is on the CATE magnitudes. On ranking the order reverses: B-SLIM has higher Spearman correlation in every cell (Table \ref{tab:simulation_spearman}). It also recovers more of the true top decile at higher precision under the interval-based selection rule (Section \ref{subsec:ranking_supp}). When the treatment function is monotone, the linear approximation therefore hurts the estimated sizes more than the ordering of patients.

\begin{table}[htbp]
\centering
\caption{Performance Comparison in the \textbf{Heavy Saturation Scenario} (Non-linear Sigmoid CATE). Results are shown for models with Heterogeneous Treatment Effects, comparing Linear and Non-linear prognostic functions ($\mu(x)$).}
\label{tab:med_sat_scenario}
\resizebox{\textwidth}{!}{
\begin{tabular}{lr ccc ccc c ccc ccc}
\toprule
& & \multicolumn{6}{c}{\textbf{BCF}} & & \multicolumn{6}{c}{\textbf{B-SLIM}} \\
\cmidrule{3-8} \cmidrule{10-15}
$\mu(x)$ Form & $N$ & \multicolumn{3}{c}{ATE} & \multicolumn{3}{c}{CATE} & & \multicolumn{3}{c}{ATE} & \multicolumn{3}{c}{CATE} \\
\cmidrule{3-5} \cmidrule{6-8} \cmidrule{10-12} \cmidrule{13-15}
& & RMSE & Cov & Len & RMSE & Cov & Len & & RMSE & Cov & Len & RMSE & Cov & Len \\
\midrule
\multirow{6}{*}{Linear}
& 250 & 0.223 & 0.96 & 0.887 & 0.608 & 0.96 & 2.559 &  & 0.217 & 0.97 & 0.894 & 0.680 & 0.91 & 2.324 \\
& 500 & 0.133 & 0.98 & 0.569 & 0.423 & 0.97 & 1.863 &  & 0.135 & 0.96 & 0.573 & 0.544 & 0.84 & 1.550 \\
& 750 & 0.110 & 0.96 & 0.447 & 0.362 & 0.96 & 1.556 &  & 0.115 & 0.96 & 0.451 & 0.515 & 0.77 & 1.265 \\
& 1000 & 0.098 & 0.96 & 0.381 & 0.323 & 0.96 & 1.376 &  & 0.099 & 0.95 & 0.384 & 0.491 & 0.70 & 1.081 \\
& 1500 & 0.077 & 0.95 & 0.306 & 0.274 & 0.96 & 1.151 &  & 0.079 & 0.94 & 0.309 & 0.472 & 0.60 & 0.877 \\
& 3000 & 0.053 & 0.96 & 0.211 & 0.211 & 0.95 & 0.857 &  & 0.054 & 0.97 & 0.214 & 0.452 & 0.42 & 0.616 \\
\midrule
\multirow{6}{*}{Non-linear}
& 250 & 0.274 & 0.92 & 0.866 & 0.710 & 0.96 & 2.731 &  & 0.238 & 0.94 & 0.842 & 0.710 & 0.90 & 2.299 \\
& 500 & 0.172 & 0.89 & 0.548 & 0.508 & 0.96 & 1.906 &  & 0.159 & 0.92 & 0.541 & 0.551 & 0.83 & 1.509 \\
& 750 & 0.115 & 0.94 & 0.431 & 0.400 & 0.96 & 1.569 &  & 0.109 & 0.95 & 0.428 & 0.500 & 0.76 & 1.193 \\
& 1000 & 0.094 & 0.96 & 0.368 & 0.347 & 0.96 & 1.380 &  & 0.089 & 0.96 & 0.365 & 0.480 & 0.69 & 1.022 \\
& 1500 & 0.082 & 0.91 & 0.292 & 0.291 & 0.95 & 1.139 &  & 0.081 & 0.92 & 0.294 & 0.464 & 0.58 & 0.817 \\
& 3000 & 0.051 & 0.95 & 0.202 & 0.210 & 0.95 & 0.836 &  & 0.051 & 0.95 & 0.204 & 0.446 & 0.39 & 0.572 \\
\bottomrule
\end{tabular}}
\end{table}

\subsubsection{Mild Saturation}

The mild saturation scenario provides a less severe departure from linearity. Table \ref{tab:mild_ate_cate} presents the RMSE, empirical coverage, and average length of the 95\% credible intervals in this setting. When the treatment effect saturates slowly as $x_1$ increases, it can be approximated by a linear gradient. From $N=500$, B-SLIM has lower CATE RMSE than BCF, with CATE coverage close to or above nominal. Its CATE intervals become shorter from $N=750$ under linear prognosis and from $N=500$ under non-linear prognosis, while at $N=250$ BCF has shorter intervals under both prognostic forms but lower CATE RMSE only under linear prognosis.

ATE coverage is close to nominal under linear prognosis, but less satisfactory in some settings under non-linear prognosis. With a linear prognostic function, the nominal $95\%$ ATE intervals have coverage ranging from $94.0\%$ to $97.5\%$ for both models across the scenarios. The favourable CATE results therefore do not imply uniformly well-calibrated inference for the ATE. Under linear prognosis, both models have mostly small positive ATE bias. In the linear treatment-effect scenario, mean ATE errors decrease from $0.016$ at $N=250$ to $0.014$ at $N=3000$ for B-SLIM, and from $0.039$ to $0.017$ for BCF. The small positive ATE bias suggests that both models may not fully adjust for selection in this setting.

\begin{table}[H]
\centering
\caption{Performance Metrics: Mild Saturation Scenario. Evaluation of the Average Treatment Effect (ATE) and Conditional Average Treatment Effect (CATE) across varying sample sizes ($N$) and functional forms of the prognostic score ($\mu(x)$).}
\label{tab:mild_ate_cate}
\resizebox{\textwidth}{!}{
\begin{tabular}{lr ccc ccc c ccc ccc}
\toprule
& & \multicolumn{6}{c}{\textbf{ BCF}} & & \multicolumn{6}{c}{\textbf{B-SLIM}} \\
\cmidrule{3-8} \cmidrule{10-15}
$\mu(x)$ Form & $N$ & \multicolumn{3}{c}{ATE} & \multicolumn{3}{c}{CATE} & & \multicolumn{3}{c}{ATE} & \multicolumn{3}{c}{CATE} \\
\cmidrule{3-5} \cmidrule{6-8} \cmidrule{10-12} \cmidrule{13-15}
& & RMSE & Cov & Len & RMSE & Cov & Len & & RMSE & Cov & Len & RMSE & Cov & Len \\
\midrule
\multirow{6}{*}{Linear}
& 250 & 0.214 & 0.96 & 0.871 & 0.465 & 0.97 & 2.032 &  & 0.217 & 0.98 & 0.877 & 0.523 & 0.97 & 2.265 \\
& 500 & 0.132 & 0.96 & 0.565 & 0.336 & 0.97 & 1.496 &  & 0.131 & 0.96 & 0.565 & 0.333 & 0.97 & 1.522 \\
& 750 & 0.108 & 0.96 & 0.445 & 0.293 & 0.96 & 1.260 &  & 0.110 & 0.95 & 0.444 & 0.280 & 0.97 & 1.238 \\
& 1000 & 0.096 & 0.96 & 0.380 & 0.268 & 0.96 & 1.121 &  & 0.096 & 0.95 & 0.378 & 0.235 & 0.97 & 1.059 \\
& 1500 & 0.077 & 0.95 & 0.305 & 0.226 & 0.96 & 0.949 &  & 0.077 & 0.95 & 0.305 & 0.192 & 0.97 & 0.858 \\
& 3000 & 0.052 & 0.97 & 0.211 & 0.174 & 0.96 & 0.715 &  & 0.052 & 0.97 & 0.212 & 0.136 & 0.97 & 0.599 \\
\midrule
\multirow{6}{*}{Non-linear}
& 250 & 0.240 & 0.93 & 0.847 & 0.566 & 0.95 & 2.086 &  & 0.233 & 0.94 & 0.828 & 0.552 & 0.95 & 2.251 \\
& 500 & 0.156 & 0.93 & 0.541 & 0.420 & 0.94 & 1.478 &  & 0.157 & 0.91 & 0.534 & 0.341 & 0.96 & 1.477 \\
& 750 & 0.108 & 0.95 & 0.427 & 0.331 & 0.95 & 1.244 &  & 0.106 & 0.95 & 0.422 & 0.252 & 0.98 & 1.167 \\
& 1000 & 0.090 & 0.95 & 0.365 & 0.296 & 0.95 & 1.117 &  & 0.088 & 0.96 & 0.360 & 0.211 & 0.98 & 0.995 \\
& 1500 & 0.081 & 0.90 & 0.291 & 0.246 & 0.96 & 0.938 &  & 0.079 & 0.91 & 0.289 & 0.170 & 0.98 & 0.791 \\
& 3000 & 0.051 & 0.96 & 0.201 & 0.176 & 0.96 & 0.695 &  & 0.049 & 0.96 & 0.200 & 0.117 & 0.98 & 0.549 \\
\bottomrule
\end{tabular}}
\end{table}

\subsubsection{Ranking Performance}

Table \ref{tab:simulation_spearman} reports Spearman correlations with the true CATE. B-SLIM has higher mean correlation in every heterogeneous simulation cell. In the linear scenario, correlations range from $0.962$ to $0.998$, compared with $0.880$ to $0.992$ for BCF. Under heavy saturation they range from $0.930$ to $0.997$, compared with $0.888$ to $0.984$ for BCF. Thus, the linear approximation can preserve the ordering of patients even when it biases the CATE magnitudes. Ranking is more difficult for both methods under mild saturation, where the treatment effect varies less.

The interval-based selection results in the Supplementary Material (Section \ref{subsec:ranking_supp}) provide a related comparison. Under heavy saturation B-SLIM recovers more of the true top decile than BCF with higher precision, despite its poor coverage in this scenario. Results are more mixed at the $25\%$ and $50\%$ thresholds. These selection results also depend on interval width and calibration, and should be read alongside the Spearman correlations. The full ranking tables, including the broader thresholds, are in the Supplementary Material.

\begin{table}[htbp]
\centering\small
\caption{Spearman correlation between posterior-mean CATE estimates and the true CATE in simulation. Values are means (Monte Carlo standard errors) over 200 replicates. Spearman correlation is undefined for the homogeneous treatment effect.}
\label{tab:simulation_spearman}
\resizebox{\textwidth}{!}{%
\begin{tabular}{llcccc}
\toprule
& & \multicolumn{2}{c}{Linear $\mu(x)$} & \multicolumn{2}{c}{Non-linear $\mu(x)$} \\
\cmidrule(lr){3-4}\cmidrule(lr){5-6}
Scenario & $N$ & BCF & B-SLIM & BCF & B-SLIM \\
\midrule
\multirow{6}{*}{Linear} & 250 & 0.902 (0.0028) & 0.966 (0.0016) & 0.880 (0.0037) & 0.962 (0.0024) \\
 & 500 & 0.955 (0.0012) & 0.985 (0.0007) & 0.943 (0.0015) & 0.985 (0.0007) \\
 & 750 & 0.970 (0.0007) & 0.990 (0.0005) & 0.965 (0.0010) & 0.992 (0.0004) \\
 & 1000 & 0.977 (0.0006) & 0.993 (0.0004) & 0.972 (0.0008) & 0.994 (0.0003) \\
 & 1500 & 0.984 (0.0003) & 0.995 (0.0002) & 0.981 (0.0005) & 0.996 (0.0002) \\
 & 3000 & 0.992 (0.0002) & 0.998 (0.0001) & 0.990 (0.0003) & 0.998 (0.0001) \\
\midrule
\multirow{6}{*}{Heavy saturation} & 250 & 0.902 (0.0031) & 0.936 (0.0033) & 0.888 (0.0043) & 0.930 (0.0049) \\
 & 500 & 0.944 (0.0019) & 0.974 (0.0013) & 0.934 (0.0020) & 0.975 (0.0013) \\
 & 750 & 0.958 (0.0014) & 0.982 (0.0009) & 0.953 (0.0015) & 0.985 (0.0008) \\
 & 1000 & 0.965 (0.0011) & 0.987 (0.0007) & 0.961 (0.0012) & 0.989 (0.0006) \\
 & 1500 & 0.973 (0.0009) & 0.991 (0.0005) & 0.972 (0.0008) & 0.993 (0.0004) \\
 & 3000 & 0.983 (0.0006) & 0.995 (0.0002) & 0.984 (0.0005) & 0.997 (0.0002) \\
\midrule
\multirow{6}{*}{Mild saturation} & 250 & 0.452 (0.0167) & 0.527 (0.0196) & 0.378 (0.0169) & 0.576 (0.0189) \\
 & 500 & 0.660 (0.0130) & 0.762 (0.0115) & 0.546 (0.0158) & 0.772 (0.0110) \\
 & 750 & 0.741 (0.0107) & 0.820 (0.0094) & 0.696 (0.0122) & 0.849 (0.0076) \\
 & 1000 & 0.788 (0.0093) & 0.858 (0.0080) & 0.752 (0.0103) & 0.892 (0.0051) \\
 & 1500 & 0.851 (0.0061) & 0.903 (0.0051) & 0.852 (0.0060) & 0.932 (0.0035) \\
 & 3000 & 0.919 (0.0033) & 0.952 (0.0024) & 0.930 (0.0027) & 0.966 (0.0016) \\
\bottomrule
\end{tabular}}
\end{table}

In a separate paired simulation of the linear interaction treatment effect, we also investigated residualising the treatment indicator. This variant increased the CATE RMSE under both prognostic forms, by $14.3$ to $36.2\%$ (Supplementary Section \ref{subsec:ortho_supp}).

Overall, BCF performs better on CATE magnitudes when the treatment effect cannot be represented by main effects and pairwise interactions. When subgroup heterogeneity is represented well by these terms, B-SLIM estimates the CATE more efficiently in these simulations.

\subsection{IHDP Results}
\label{subsec:ihdp}

The IHDP benchmark is a challenging setting for B-SLIM because the true treatment effect is the difference between linear and exponential mean outcome surfaces. Despite this structural mismatch, the model approximates the CATE surface and performs on par with BCF (Table \ref{tab:ihdp_primary}). BCF has a slightly lower mean RMSE ($1.720$ versus $1.739$ for B-SLIM), a lower median RMSE, and the lower RMSE in 75 of the 100 data sets. BCF also has the lower mean rank and a calibration slope closer to one, as well as a higher mean Spearman correlation ($0.892$ versus $0.879$).

Of the five implementations, the two Bayesian methods have the lowest mean RMSE and the lowest mean rank, and the two highest mean Spearman correlations. Calibration is more mixed: BCF is closest to a slope of one. Both models use the same BART prior for the prognostic function and differ on the priors chosen to model the treatment contrast. We observe that the pairwise interactions still approximate the non-linear treatment function over the observed covariate support, because they allow the effect of one covariate to depend on another.

The signal-stratified results in Table \ref{tab:ihdp_signal_bands} reverse the ordering of the two Bayesian methods. BCF has the lower RMSE in 24 of the 25 data sets in the weakest quartile, whereas B-SLIM has the lower RMSE in 16 of the 25 data sets in the strongest quartile. This matches the simulations. A linear treatment function uses a structured signal more efficiently, while BCF shrinks more towards a homogeneity. When the treatment effect varies less, as in the weakest quartile, the extra shrinkage helps. When the CATE signal is pronounced, B-SLIM fares better. The quartiles also differ in the scale and shape of $\tau$, so signal strength is probably not the only reason. What the split does show is that the two Bayesian models do not rank the same way in every IHDP scenario.

\begin{table}[H]
\centering\footnotesize
\caption{IHDP benchmark. Results over 100 replicates using 10-fold out-of-sample predictions. Values are mean (standard error) unless stated. RMSE is for the CATE. The mean rank is the average rank by RMSE across the 100 data sets.}
\label{tab:ihdp_primary}
\resizebox{\textwidth}{!}{%
\begin{tabular}{lccccc}
\toprule
Method & RMSE & median RMSE & Spearman & Calib.\ slope & mean rank\\
\midrule
B-SLIM & 1.739 (0.242) & 0.957 & 0.879 (0.015) & 0.808 (0.022) & 2.08\\
BCF & 1.720 (0.269) & 0.819 & 0.892 (0.018) & 0.928 (0.022) & 1.36\\
DR-learner & 2.343 (0.250) & 1.618 & 0.716 (0.029) & 0.786 (0.033) & 3.88\\
S-learner & 3.109 (0.453) & 1.545 & 0.817 (0.023) & 1.822 (0.085) & 3.41\\
R-learner & 4.212 (0.635) & 1.912 & 0.726 (0.017) & 1.715 (0.058) & 4.27\\
\bottomrule
\end{tabular}
}
\end{table}

\begin{table}[H]
\centering\small
\caption{Performance of the two Bayesian methods by the amount of treatment
effect signal present, measured by quartiles of $\operatorname{sd}(\tau)$ across
the 100 data sets. $\Delta$ is the difference in RMSE (for the CATE) between B-SLIM and BCF, so a negative value favours B-SLIM.}
\label{tab:ihdp_signal_bands}
\begin{tabular}{ccrrrr}
\toprule
Quartile & range of $\operatorname{sd}(\tau)$ & data sets & mean $\Delta$ & median $\Delta$ & B-SLIM\ wins\\
\midrule
1 (weakest) & 0.03--1.08 & 25 & 0.139 & 0.135 & 1/25\\
2 & 1.15--2.54 & 25 & 0.126 & 0.123 & 2/25\\
3 & 2.57--5.55 & 25 & 0.149 & 0.162 & 6/25\\
4 (strongest) & 5.67--43.22 & 25 & $-0.338$ & $-0.228$ & 16/25\\
\bottomrule
\end{tabular}
\end{table}

\subsection{Individual CATE Posteriors on IHDP}
\label{subsec:ihdp_posterior}

In addition to the point estimates, we also evaluate whether the models correctly predict harm or benefit, and how they quantify uncertainty. The IHDP outcomes are synthetic, so we can check the sign of each posterior and whether the interval covers the true effect. We report how often the $95\%$ interval excludes zero, and the Type S error among those that do (Table \ref{tab:ihdp_posterior}). The same quantities are split by quartiles of signal strength of the treatment effect in Table \ref{tab:ihdp_posterior_bands}.

The main difference is the length of the credible intervals. B-SLIM has wider intervals, while the shorter intervals of BCF exclude zero more often. This extra confidence comes at a price. BCF undercovers more than B-SLIM in every signal quartile, and its Type S rate is slightly higher overall, although not in the strongest quartile (Table \ref{tab:ihdp_posterior_bands}). The AUC for $P(\tau_i>0)$ as a classifier of the sign of $\tau_i$, i.e. how well the model can discriminate between a true negative and true positive effect is higher for B-SLIM. Coverage is better and more uniform for B-SLIM, but still below the nominal $95\%$ for both methods except in the weakest quartile.

The exclusion rates before calibration in Table~\ref{tab:ihdp_posterior} are therefore not a fair comparison of how often each model can detect a non-zero effect. The B-SLIM intervals cover $93.0\%$ of the true $\tau_i$, against $91.4\%$ for BCF. We use the posterior draws to rescale each model's intervals towards a common empirical coverage target of $95\%$. The amount of adjustment was chosen from the remaining 99 data sets and then applied to the held-out one, so the same replicate is not used both to set the interval and to score it. The Supplementary Material (Section~\ref{subsec:ihdp_matched_supp}, Table~\ref{tab:ihdp_posterior_matched}) reports the result. BCF's intervals have to be expanded the most ($c=1.186$), while B-SLIM needs $c=1.105$. At approximately matched coverage, the exclusion rates are close: $87.6\%$ of children for B-SLIM and $87.8\%$ for BCF (paired difference $-0.0015$, $95\%$ CI $[-0.0101,0.0072]$). The higher exclusion rate of BCF before calibration therefore becomes much smaller at comparable empirical coverage.

\begin{table}[H]
\centering\footnotesize
\caption{Individual CATE posteriors on IHDP (100 replicates, 10-fold out-of-sample,
$95\%$ credible intervals). Values are mean (standard error) over the 100 data sets, except AUC, which is averaged where both true effect signs occur.
``Excludes $0$'' is the fraction of children whose interval excludes zero.
``Sign correct'' and ``Type S'' are conditional on that. Coverage is of the true $\tau_i$
(nominal $0.95$). AUC is for $P(\tau_i>0)$ as a classifier of $\operatorname{sign}(\tau_i)$.}
\label{tab:ihdp_posterior}
\resizebox{\textwidth}{!}{%
\begin{tabular}{lcccccc}
\toprule
Method & excludes $0$ & sign correct & Type S & coverage & width & AUC $P(\tau>0)$\\
\midrule
\textbf{B-SLIM} & 0.890 (0.007) & \textbf{0.998} (0.000) & \textbf{0.002} (0.000) & \textbf{0.930} (0.005) & 5.04 (0.55) & \textbf{0.992} (0.001)\\
BCF & \textbf{0.897} (0.009) & 0.997 (0.000) & 0.003 (0.000) & 0.914 (0.005) & \textbf{4.55} (0.64) & 0.976 (0.008)\\
\bottomrule
\end{tabular}}
\end{table}

\begin{table}[H]
\centering\small
\caption{Coverage of the true $\tau_i$ and Type S error rate by quartile of
$\operatorname{sd}(\tau)$, using the same quartiles as Table
\ref{tab:ihdp_signal_bands}. Nominal coverage is $0.95$.}
\label{tab:ihdp_posterior_bands}
\begin{tabular}{lcccc}
\toprule
& \multicolumn{2}{c}{Coverage} & \multicolumn{2}{c}{Type S (\%)}\\
\cmidrule(lr){2-3}\cmidrule(lr){4-5}
Quartile of $\operatorname{sd}(\tau)$ & BCF & B-SLIM & BCF & B-SLIM\\
\midrule
1 (weakest) & 0.961 & \textbf{0.963} & 0.05 & \textbf{0.02}\\
2 & 0.928 & \textbf{0.944} & 0.38 & \textbf{0.11}\\
3 & 0.898 & \textbf{0.932} & 0.29 & \textbf{0.17}\\
4 (strongest) & 0.870 & \textbf{0.880} & \textbf{0.41} & 0.47\\
\bottomrule
\end{tabular}
\end{table}

\section{Applied Example: The CANVAS Program}
\label{subsec:applied_example}
We applied B-SLIM to individual participant data from the CANVAS Program. We consider two randomized, double-blind, placebo-controlled trials, CANVAS (NCT01032629) and CANVAS-R (NCT01989754), designed to evaluate the cardiovascular and renal effects of the SGLT2 inhibitor canagliflozin in patients with type 2 diabetes at high cardiovascular risk.\citep{neal2013rationale, neal2017rationale} Access to this trial data was granted under the Yale University Open Data Access (YODA) Project (Access Request Number: 2025-0684 \citep{yodacanvasdata}). The primary continuous outcome measure for this analysis is the change in HbA1c from baseline to the 52-week follow-up. HbA1c reflects average blood glucose over the preceding months and is an important measure of the glucose-lowering response to SGLT2 inhibitors. Earlier studies of canagliflozin have also assessed glycaemic efficacy over this follow-up period.\citep{cefalu2013efficacy}

We compare canagliflozin with placebo. To capture potential drivers of treatment effect heterogeneity, we selected a set of baseline participant characteristics, including baseline HbA1c, age, urinary albumin-to-creatinine ratio, eGFR, systolic blood pressure, history of cerebrovascular disease, diabetes duration, and sex. We also included body mass index, recorded race, history of heart failure, and baseline use of insulin, metformin, GLP-1 receptor agonists, and agents acting on the renin--angiotensin system. The full covariate list and coding are given in Supplementary Section \ref{SEC:supplement_canvas}. We used complete cases for the 52-week HbA1c change, treatment assignment, and selected baseline covariates; missing values were not imputed. B-SLIM uses symmetric treatment coding to give both mean outcome functions the same prior variability, while BCF uses adaptive treatment coding. Both models include the treatment proportion in the full CANVAS-R training pool as the same constant propensity score in the prognostic component. The Shapley decomposition and coefficient posteriors are presented below.

\subsection{Results and Interpretability}
\label{subsec:oos_performance_summary}

We trained both models on the CANVAS-R analytic sample and evaluated them in the CANVAS analytic sample. In the smaller training samples, both models shrink the estimated treatment effects towards homogeneity, more so for BCF (Supplementary Section \ref{SEC:supplement_canvas}). At the full CANVAS-R sample, the correlation between their out-of-sample CATE predictions is $r=0.843$; at $N=4000$ it is $0.813$ (Table \ref{tab:model_metrics}). The models rank patients similarly by predicted benefit, while B-SLIM has lower out-of-sample RMSE for HbA1c change at every training size. When their rankings are similar, B-SLIM also gives direct parameter-level inference on the treatment-effect modifiers.

\begin{figure}[p]
    \centering
    \includegraphics[width=\textwidth]{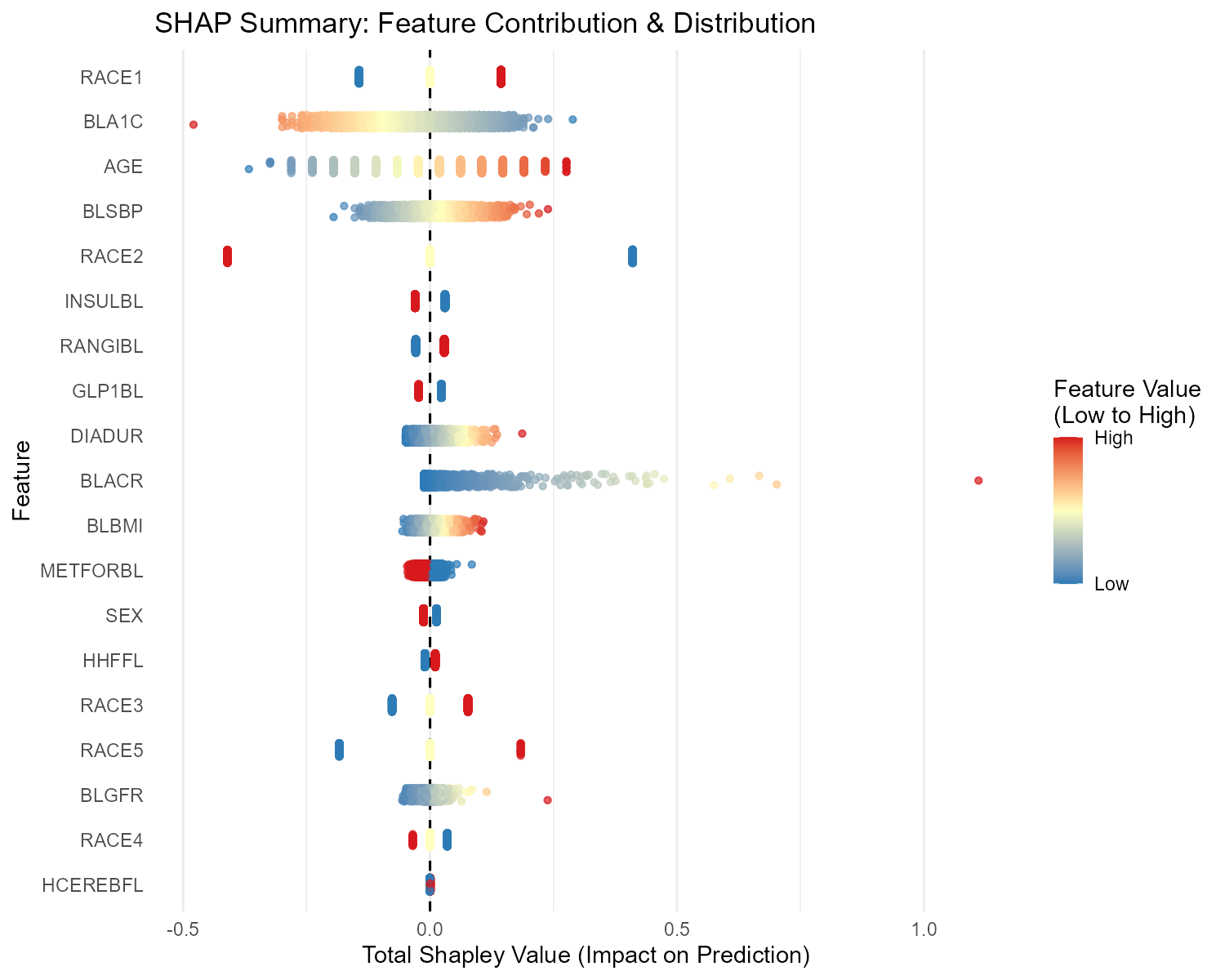}
    \caption{Individual Shapley summaries for the predicted CATE in CANVAS-R. Each point adds the posterior medians of one individual's main-effect and interaction contributions. Colour indicates the covariate value, from low (blue) to high (red). Negative contributions indicate a larger HbA1c reduction. RACE1--RACE5 are contrast columns of one categorical covariate. The spread describes variation between individuals. Abbreviations are defined in Supplementary Table \ref{tab:canvas_covariates}.}
    \label{fig:shapley_contribution}
\end{figure}

RACE, considered jointly, has the largest grouped Shapley contribution, followed by baseline HbA1c and age; these contributions arise mainly through main effects. Figure \ref{fig:shapley_contribution} shows their contributions across individuals. Higher baseline HbA1c generally contributes to a larger treatment-related reduction in HbA1c, while older age contributes to a smaller reduction.

We constructed simultaneous $95\%$ posterior intervals across the full set of assessed treatment-effect contrasts for all covariates, including all pairwise RACE comparisons. The simultaneous intervals for baseline HbA1c, age, and the Black or African American versus White contrast exclude zero. For the RACE contrast, the estimated treatment-related reduction in HbA1c is greater in Black or African American participants than in White participants (Supplementary Figure \ref{fig:canvas_full_contrasts}). These contrasts account for interactions and are averaged over the observed distribution of the remaining covariates (Supplementary Section \ref{subsec:applied_interpretability}). BCF provides posterior distributions of the CATE, but does not offer the same direct inference on individual effect modifiers. A pooled analysis that included CANVAS sub-studies reported HbA1c reductions across racial groups, although the studies were not powered for comparisons between them.\citep{gavin2015racial} Another CANVAS Program analysis compared cardiovascular and renal outcomes in Asian and White participants,\citep{chen2024ethnic} although they considered different outcomes than our analysis.

\subsection{Posterior Distributions of the Main Effects}
\label{sec:posterior_behaviour}
\begin{figure}[H]
    \centering
    \includegraphics[width=0.60\linewidth]{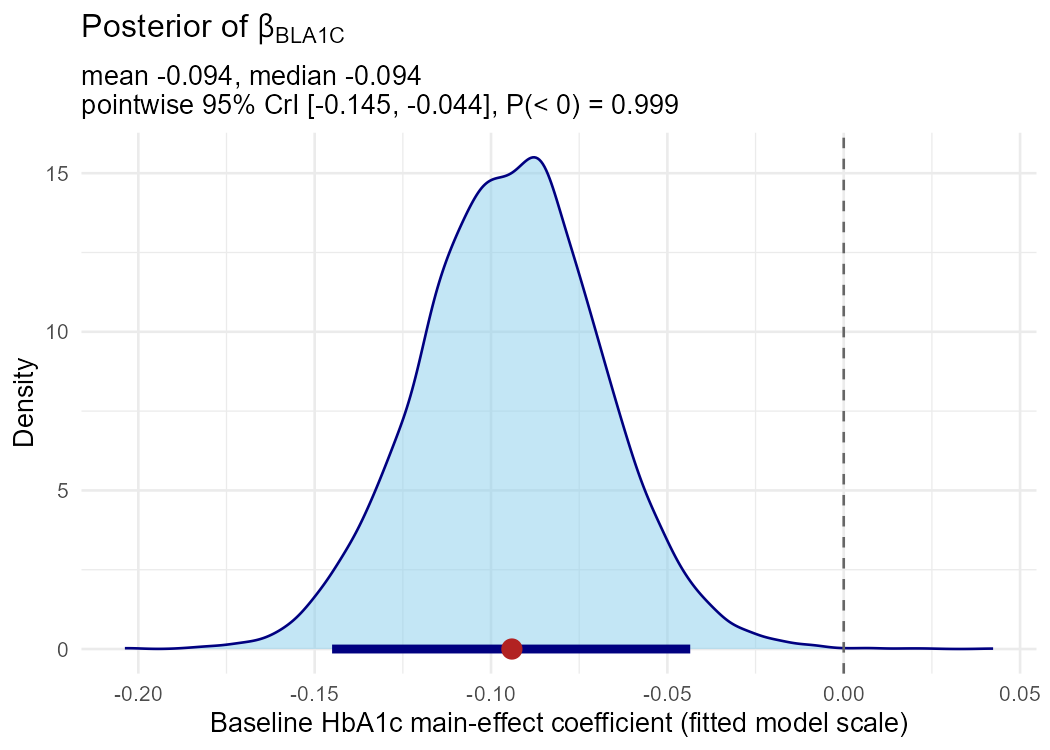}
    \par\medskip
    \includegraphics[width=0.60\linewidth]{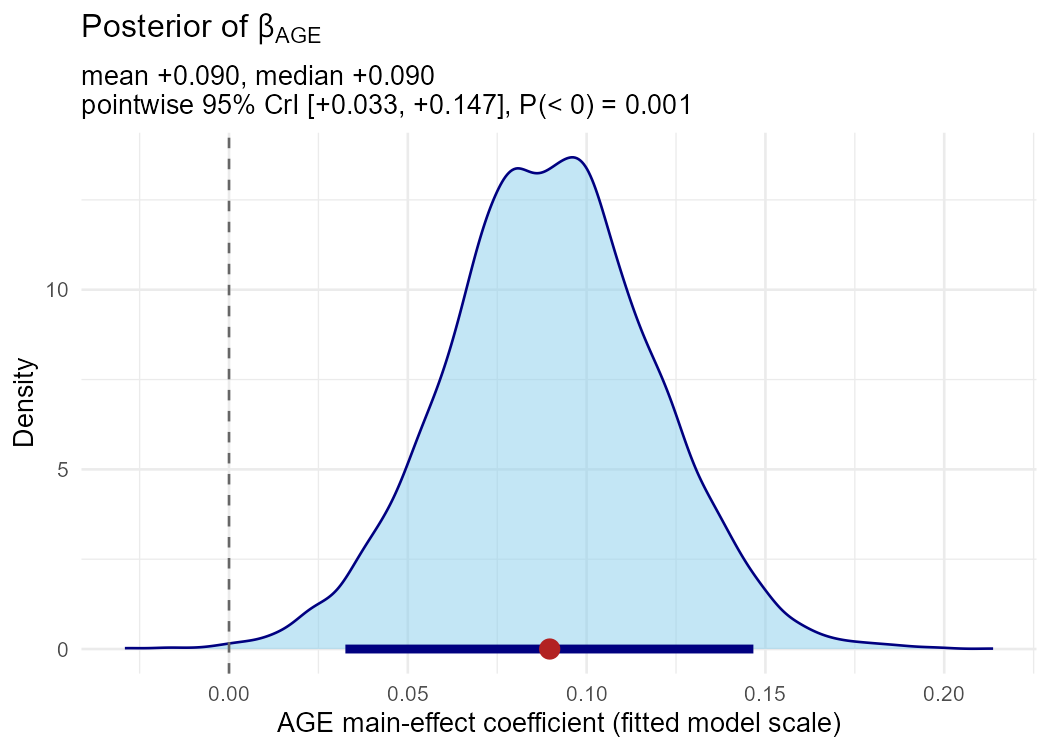}
    \caption{Posterior distributions of the main-effect coefficients for baseline HbA1c (upper panel) and age (lower panel). Both pointwise $95\%$ credible intervals exclude zero.}
    \label{fig:posterior_HBA1C}
    \label{fig:posterior_AGE}
\end{figure}

Figure \ref{fig:posterior_HBA1C} shows the posteriors of the main-effect coefficients for baseline HbA1c and age. Their means are $-0.094$ and $0.090$, with pointwise $95\%$ credible intervals $[-0.145,-0.044]$ and $[0.033,0.147]$, respectively, on the fitted model scale. With all other encoded covariates set to zero, the main-effect coefficients indicate that higher baseline HbA1c is associated with a greater treatment-related reduction in HbA1c, and older age with a smaller reduction.

\FloatBarrier

\section{Code and Data Availability}
\label{sec:availability}
The simulation, IHDP, and application analysis scripts are available at \url{https://github.com/HYHBalci/bslim-paper}.\citep{bslimpaper} The B-SLIM implementation for general use is available in \texttt{bslim}, based on \texttt{stochtree}, at \url{https://github.com/HYHBalci/bslim}.\citep{bslimsoftware} The analysis repository includes data-generating code, model fitting, and evaluation scripts. CANVAS participant data are available through the Yale University Open Data Access Project under \href{https://yoda.yale.edu/data-request/2025-0684/}{request 2025-0684}, subject to its access requirements.

\section{Discussion}
\label{sec:discussion}
We presented B-SLIM, which places a flexible BART prior on the prognostic function and uses a sparse linear model, with main effects and pairwise interactions, to capture the treatment effect. The global--local shrinkage prior enables the search of a large interaction space and allows appropriate inference. Shapley values provide interpretation by summarizing how each covariate contributes to the CATE. The model also provides conditional mean potential-outcome predictions, besides CATE estimation.

The simulations describe the trade-off that our semi-parametric structure creates. When the treatment effect is linear, or close enough to a linear gradient, B-SLIM generally has lower CATE RMSE and, at larger sample sizes, tighter credible intervals than BCF. When the treatment function is more strongly non-linear, BCF has lower RMSE and better coverage. Spearman ranking favours B-SLIM in all heterogeneous simulation settings. Under monotone saturation it is also generally the more sensitive and the more precise of the two at the sharpest interval-based threshold. ATE coverage is close to nominal under linear prognosis: both models' nominal $95\%$ ATE intervals have coverage between $94.0\%$ and $97.5\%$, alongside mostly small positive ATE bias. Residual bias under targeted selection may contribute, although these results do not identify its source.

We primarily compare with BCF because both models use a BART prognostic component, allowing us to examine the effect of modelling the treatment contrast parametrically or with a second forest. IHDP provides a broader comparison with the DR-, S-, and R-learners.

On IHDP the true treatment effect is the difference between linear and exponential mean outcome surfaces, so it does not lie in the linear-plus-interactions span of our model. Even so, the approximation is close enough that B-SLIM performs on par with BCF. The two Bayesian methods have the first and second lowest mean RMSE and mean rank, they have the two highest mean Spearman correlations with the true CATE, and both outperform the implementations of the frequentist machine learners considered here (the DR-learner, the S-learner, and the R-learner) on these metrics. Comparing the two Bayesian methods on individual inference, BCF excludes zero from the individual intervals more often, while B-SLIM has wider intervals, better coverage of the true $\tau_i$, and fewer Type S errors overall. Neither method attains the nominal $95\%$ coverage except in the weakest quartile. After calibrating the intervals towards the same empirical coverage, B-SLIM and BCF have very similar exclusion rates.

In our data set CANVAS, the two models rank patients similarly, while B-SLIM has lower out-of-sample outcome prediction error. Its grouped Shapley decomposition places RACE, baseline HbA1c, and age among the largest contributors to treatment-effect heterogeneity, mainly through main effects. The corresponding full CATE contrasts have simultaneous $95\%$ posterior intervals that exclude zero. B-SLIM provides these contrasts directly; BCF's CATE posterior does not give the same parameter-level inference.

The gain in efficiency of B-SLIM still comes with a structural limitation: non-linearity in the treatment function cannot always be represented by main effects and pairwise interactions. The IHDP results qualify this limitation, since a smooth non-linear treatment function can still be approximated using the main and interaction effects. The heavy saturation results show where this approximation starts to break down.

Including more covariates and their pairwise interactions increases the number of coefficients and the computation required for fitting.

\subsection{Guidance for Practice}
In practice, B-SLIM is the more natural choice when treatment effect heterogeneity is expected to be represented well by main effects and pairwise interactions, and when direct inference on effect modifiers or prediction of both potential outcomes is required. BCF is safer when strong non-linearity in the treatment function is plausible.

In real data, where the CATE is not known, out-of-sample outcome prediction can help to decide between the two models. In a randomized validation sample, patients can also be grouped by predicted benefit. Comparing the observed treatment--control differences with the predicted mean treatment effects in these groups then assesses calibration at group level. When performance is similar, we prefer B-SLIM, because it provides more interpretability at the parameter level. We also advise evaluating the agreement between the CATE rankings of the two models, because this assesses the robustness of the ranking with respect to model choice. Strong disagreement between the models calls for further model checking rather than favouring either model by itself.

\subsection{Future Directions}
It may be worthwhile to extend B-SLIM with the BART updates proposed by Prado et~al.\citep{prado2025shared} to improve identifiability. Note, however, that in our setting identifiability is much less of an issue than in the general case, as the two parts of the model act on different outcome spaces: one on the overall prediction and the other on the treatment contrast. Moreover, our sparse regularization on the regression parameters likely also helps to disentangle tree effects from regression effects. Still, more targeted updating may improve convergence of the parameter estimates and lower their uncertainty.

Rather than estimating a sparse signal with a treatment function, future work might directly classify whether individuals do or do not benefit from a treatment. Recent work by M{\"u}ller et~al.\citep{muller2025data} and Cheng et~al.\citep{cheng2025chiseling} provides frameworks for identifying subgroups with treatment benefit while controlling specified error rates under their respective assumptions. Given that our model can capture the prognostic function, the remaining partial residuals could be used to classify patients into subgroups without explicitly modelling the treatment effect function. Furthermore, Chen et~al.\citep{chen2026comparing} present a global test for heterogeneity based on whether score residuals are associated with the covariates. A formulation that uses a BART prior for the prognostic function could extend this test. The same backfitting scheme also extends to a Student-$t$ working likelihood for continuous outcomes with heavy tails, and to a probit model for binary outcomes; both variants are implemented, but further testing is required to evaluate these. When the coefficient space grows, prior specifications that link shrinkage across main and interaction effects may improve estimation.\citep{van2024linked} Future work could also examine how more flexible propensity-score models account for the targeted-selection mechanism.

\section{Conclusion}
\label{sec:conclusion}
In summary, we provide B-SLIM, a semi-parametric Bayesian model for individualized treatment effect estimation in observational studies and randomized trials. It combines a non-parametric prognostic component with a sparse linear treatment effect component, providing direct parameter inference and patient-specific predictions of conditional mean potential outcomes. In the simulations and IHDP, the interaction basis is efficient under linearity and also approximates smooth non-linearity over the observed covariate support. On IHDP, both Bayesian methods outperform the DR-learner, S-learner, and R-learner implementations considered here in mean CATE RMSE and ranking. In CANVAS, the two Bayesian models rank patients similarly, while B-SLIM provides evidence of effect modification by baseline HbA1c, age, and RACE. B-SLIM is the more suitable choice when main effects and pairwise interactions can reasonably describe or approximate treatment effect heterogeneity and inference on the effect modifiers is of interest. BCF remains the safer default when the treatment effect is strongly non-linear and cannot be adequately represented by this basis. If the two models perform similarly, we prefer B-SLIM for its interpretability.

\section*{Acknowledgements}
This study, carried out under YODA Project \# 2025-0684, used data obtained from the Yale University Open Data Access Project, which has an agreement with JANSSEN RESEARCH \& DEVELOPMENT, L.L.C.. The interpretation and reporting of research using this data are solely the responsibility of the authors and does not necessarily represent the official views of the Yale University Open Data Access Project or JANSSEN RESEARCH \& DEVELOPMENT, L.L.C.. The original proposal can be found: \url{https://yoda.yale.edu/data-request/2025-0684/}.

\section*{Funding}
This work was supported by a starter grant awarded to Jeroen Hoogland (project number 20113364).

\begin{samepage}
\section*{Supporting information}
The accompanying supplement contains the treatment-residualisation comparison, additional simulation results, coverage-matched IHDP analyses, CANVAS validation and interpretability details, and posterior sampling equations and algorithm.
\par\end{samepage}

\clearpage
\begingroup
\setcounter{section}{0}
\setcounter{subsection}{0}
\setcounter{subsubsection}{0}
\setcounter{table}{0}
\setcounter{figure}{0}
\setcounter{equation}{0}
\setcounter{algorithm}{0}
\renewcommand{\thesection}{S\arabic{section}}
\renewcommand{\thetable}{S\arabic{table}}
\renewcommand{\thefigure}{S\arabic{figure}}
\renewcommand{\theequation}{S\arabic{equation}}
\renewcommand{\thealgorithm}{S\arabic{algorithm}}
\renewcommand{\theHsection}{supp.\arabic{section}}
\renewcommand{\theHtable}{supp.\arabic{table}}
\renewcommand{\theHfigure}{supp.\arabic{figure}}
\renewcommand{\theHequation}{supp.\arabic{equation}}
\providecommand{\theHalgorithm}{}
\renewcommand{\theHalgorithm}{supp.\arabic{algorithm}}
\pdfbookmark[0]{Supplementary Material}{supplement}
\section*{Supplementary Material}
\label{SEC:supplement}
\noindent This supplement provides additional simulation and application results and computational details.

\section{Orthogonalisation of the Treatment Contrast}
\label{subsec:ortho_supp}

The standard fit uses centered treatment coding, $z-0.5$. In the orthogonalised fit, we use $z-\hat\pi(x)$ instead. We estimate $\hat\pi(x)$ once with a Bayesian logistic regression including main effects and pairwise interactions, and hold it fixed across chains. Both fits use the same prognostic forest specification and prior. Residualising treatment changes the conditional outcome function represented by the prognostic component.

In this comparison, orthogonalising the treatment contrast increases CATE RMSE in every design, by $14.3$ to $36.2\%$ (Table~\ref{tab:ortho}). Under a linear $\mu$, the increase averages $17.7\%$, with wider intervals and higher coverage. Under a non-linear $\mu$, the increase averages $29.6\%$ and average coverage falls despite wider intervals. The last two columns of Table~\ref{tab:ortho} use the same interval-based top-quartile rule as in the main simulations; sensitivity falls in both panels.

The standard-arm values come from this separate experiment and differ from those in main-text Table \ref{tab:default_scenario}. There are 200 replicates at each sample size.

\begin{table}[htbp]
\centering\small
\caption{Orthogonalised versus standard B-SLIM under both prognostic forms, with 200 replicates per setting. Change is the percentage increase in CATE RMSE. Sensitivity uses the same interval-based top-quartile rule as in the main simulations.}
\label{tab:ortho}
\resizebox{\textwidth}{!}{%
\begin{tabular}{rccccccccc}
\toprule
& \multicolumn{3}{c}{CATE RMSE} & \multicolumn{2}{c}{Coverage} & \multicolumn{2}{c}{CI width} & \multicolumn{2}{c}{Top-25\% sensitivity}\\
\cmidrule(lr){2-4}\cmidrule(lr){5-6}\cmidrule(lr){7-8}\cmidrule(lr){9-10}
$N$ & standard & residualised & change & standard & residualised & standard & residualised & standard & residualised\\
\midrule
\multicolumn{10}{l}{\emph{Linear $\mu$}}\\
250 & 0.639 & 0.751 & $+17.4\%$ & 0.942 & 0.981 & 2.415 & 3.516 & 0.491 & 0.298 \\
500 & 0.381 & 0.459 & $+20.5\%$ & 0.959 & 0.977 & 1.619 & 2.102 & 0.621 & 0.490 \\
750 & 0.312 & 0.357 & $+14.3\%$ & 0.963 & 0.977 & 1.307 & 1.616 & 0.689 & 0.586 \\
1000 & 0.260 & 0.299 & $+14.9\%$ & 0.965 & 0.976 & 1.116 & 1.363 & 0.732 & 0.649 \\
1500 & 0.205 & 0.245 & $+19.4\%$ & 0.969 & 0.975 & 0.900 & 1.088 & 0.776 & 0.714 \\
3000 & 0.143 & 0.170 & $+19.4\%$ & 0.967 & 0.969 & 0.623 & 0.742 & 0.843 & 0.808 \\
\midrule
\multicolumn{10}{l}{\emph{Non-linear $\mu$}}\\
250 & 0.578 & 0.748 & $+29.6\%$ & 0.957 & 0.960 & 2.318 & 3.132 & 0.465 & 0.373 \\
500 & 0.350 & 0.477 & $+36.2\%$ & 0.968 & 0.961 & 1.532 & 1.945 & 0.612 & 0.548 \\
750 & 0.266 & 0.347 & $+30.1\%$ & 0.977 & 0.972 & 1.219 & 1.497 & 0.676 & 0.616 \\
1000 & 0.223 & 0.296 & $+32.8\%$ & 0.978 & 0.970 & 1.037 & 1.268 & 0.717 & 0.651 \\
1500 & 0.176 & 0.223 & $+26.7\%$ & 0.978 & 0.975 & 0.828 & 0.998 & 0.776 & 0.732 \\
3000 & 0.123 & 0.151 & $+22.4\%$ & 0.979 & 0.977 & 0.574 & 0.681 & 0.841 & 0.808 \\
\bottomrule
\end{tabular}}
\end{table}

\section{Homogeneous Simulation Results}
Table \ref{tab:hom_scenario} presents the results for the homogeneous simulation scenario, where $\tau(x)=3$ and all non-intercept treatment-effect coefficients are zero. The results show that B-SLIM and BCF can shrink back towards homogeneity, with lower CATE RMSE for BCF under linear prognosis.

\begin{table}[htbp]
\centering
\caption{Performance Comparison in the \textbf{Homogeneous Scenario} (Constant CATE, $\tau=3$). Results are shown for models with Homogeneous Treatment Effects, comparing Linear and Non-linear prognostic functions ($\mu$).}
\label{tab:hom_scenario}
\resizebox{\textwidth}{!}{
\begin{tabular}{lr ccc ccc c ccc ccc}
\toprule
& & \multicolumn{6}{c}{\textbf{BCF}} & & \multicolumn{6}{c}{\textbf{B-SLIM}} \\
\cmidrule{3-8} \cmidrule{10-15}
$\mu(x)$ Form & $N$ & \multicolumn{3}{c}{ATE} & \multicolumn{3}{c}{CATE} & & \multicolumn{3}{c}{ATE} & \multicolumn{3}{c}{CATE} \\
\cmidrule{3-5} \cmidrule{6-8} \cmidrule{10-12} \cmidrule{13-15}
& & RMSE & Cov & Len & RMSE & Cov & Len & & RMSE & Cov & Len & RMSE & Cov & Len \\
\midrule
\multirow{6}{*}{Linear}
& 250 & 0.215 & 0.97 & 0.869 & 0.388 & 0.99 & 1.956 &  & 0.217 & 0.97 & 0.875 & 0.502 & 0.97 & 2.241 \\
& 500 & 0.132 & 0.95 & 0.562 & 0.257 & 0.98 & 1.360 &  & 0.132 & 0.97 & 0.563 & 0.321 & 0.98 & 1.503 \\
& 750 & 0.108 & 0.96 & 0.444 & 0.212 & 0.99 & 1.110 &  & 0.111 & 0.95 & 0.445 & 0.269 & 0.98 & 1.225 \\
& 1000 & 0.096 & 0.96 & 0.378 & 0.191 & 0.99 & 0.967 &  & 0.097 & 0.96 & 0.378 & 0.226 & 0.98 & 1.047 \\
& 1500 & 0.076 & 0.95 & 0.304 & 0.157 & 0.98 & 0.788 &  & 0.077 & 0.94 & 0.305 & 0.186 & 0.98 & 0.850 \\
& 3000 & 0.052 & 0.97 & 0.211 & 0.109 & 0.98 & 0.554 &  & 0.052 & 0.97 & 0.211 & 0.132 & 0.97 & 0.594 \\
\midrule
\multirow{6}{*}{Non-linear}
& 250 & 0.236 & 0.92 & 0.837 & 0.474 & 0.97 & 1.955 &  & 0.234 & 0.93 & 0.824 & 0.536 & 0.96 & 2.221 \\
& 500 & 0.153 & 0.92 & 0.536 & 0.307 & 0.97 & 1.311 &  & 0.157 & 0.91 & 0.533 & 0.330 & 0.97 & 1.459 \\
& 750 & 0.108 & 0.95 & 0.425 & 0.232 & 0.98 & 1.070 &  & 0.107 & 0.95 & 0.421 & 0.245 & 0.98 & 1.152 \\
& 1000 & 0.088 & 0.97 & 0.363 & 0.205 & 0.98 & 0.937 &  & 0.089 & 0.97 & 0.359 & 0.204 & 0.98 & 0.983 \\
& 1500 & 0.080 & 0.92 & 0.290 & 0.174 & 0.98 & 0.760 &  & 0.080 & 0.91 & 0.288 & 0.166 & 0.98 & 0.785 \\
& 3000 & 0.050 & 0.97 & 0.201 & 0.122 & 0.99 & 0.527 &  & 0.049 & 0.96 & 0.200 & 0.114 & 0.98 & 0.545 \\
\bottomrule
\end{tabular}}
\end{table}

\section{Coverage-matched Individual CATE Inference\\on IHDP}
\label{subsec:ihdp_matched_supp}

This calibration is possible because the individual treatment effects are known in the semi-synthetic benchmark.

The original $95\%$ credible intervals of the two Bayesian methods do not have the same coverage of the true $\tau_i$ on IHDP, so their exclusion rates are not directly comparable. Let $m_i$ be the posterior mean, and $q_{i,0.025}$ and $q_{i,0.975}$ the posterior percentiles that form the central $95\%$ credible interval. We rescale this interval as
\[
 I_i(c)=\left[m_i+c\{q_{i,0.025}-m_i\},\;m_i+c\{q_{i,0.975}-m_i\}\right].
\]
The scale factor $c>1$ widens the interval. For each held-out replicate and each model, we choose the smallest $c$ attaining at least $95\%$ empirical coverage across the other 99 data sets, and then evaluate it on the held-out replicate. The resulting intervals are empirically calibrated CATE intervals, not tolerance intervals or necessarily $95\%$ posterior credible intervals. The shared covariates and treatment assignments mean this evaluates calibration across IHDP response surfaces rather than across new patient samples. Table~\ref{tab:ihdp_posterior_matched} reports the comparison; held-out mean coverage is $0.950$ for both methods. The exclusion rates are close.

\begin{table}[htbp]
\centering\footnotesize
\caption{Coverage-matched individual CATE inference on IHDP. Each model's
intervals are rescaled towards a common empirical coverage target of $95\%$. The scale
factor is chosen from the other 99 data sets and then applied to the held-out
one; held-out coverage need not equal the target exactly. Values are mean (standard error) over 100 replicates. The scale factor
$c$ multiplies the distances from the posterior mean to the $2.5$th and $97.5$th posterior percentiles; $c>1$ widens the interval. ``Correct exclusion'' is the fraction
of children whose interval excludes zero in the direction of the true effect.}
\label{tab:ihdp_posterior_matched}
\resizebox{\textwidth}{!}{%
\begin{tabular}{lcccccc}
\toprule
Method & $c$ & coverage & excludes $0$ & correct exclusion & Type S & width\\
\midrule
\textbf{B-SLIM} & 1.105 (0.000) & 0.950 (0.004) & 0.876 (0.008) & 0.875 (0.008) & 0.001 (0.000) & 5.56 (0.61)\\
BCF & 1.186 (0.000) & 0.950 (0.003) & 0.878 (0.010) & 0.876 (0.010) & 0.002 (0.000) & 5.39 (0.76)\\
\bottomrule
\end{tabular}}
\end{table}

\section{Detailed Analysis: The CANVAS Program}
\label{SEC:supplement_canvas}

\subsection{Out-of-Sample Performance}
\label{subsec:oos_performance}

We compare B-SLIM with BCF by using an analytic sample of 5018 individuals from CANVAS-R for training and 3004 individuals from CANVAS for validation. These are the numbers included in this analysis, rather than the original trial enrolments. To study the effect of sample size, we drew 20 treatment-stratified samples without replacement for $N \in \{500, 1000, 1500, 2000, 3000, 4000\}$ from CANVAS-R and trained both models on each sample. We also fit both models once on the full CANVAS-R sample. Both models include the treatment proportion in the full CANVAS-R training pool as the same constant propensity score in the prognostic component. We evaluate agreement between the models using the correlation between their out-of-sample CATEs, and we evaluate ranking using the uplift curve.\citep{rzepakowski2010decision} Because the outcome is follow-up minus baseline HbA1c, a more negative CATE indicates greater benefit. The uplift curve ranks individuals from high to low predicted benefit, $-\hat\tau(x)$, and plots the cumulative gain as an increasing portion of the population is treated. This evaluation relies on randomization; without exchangeability, the empirical treatment effect cannot be compared directly with the model output. Figure \ref{fig:cv_dashboards} and Table \ref{tab:model_metrics} show the results for both methods at the different sample sizes.

The comparison of the individual CATEs shows that the two models already agree reasonably well at small $N$, and that this agreement increases with sample size (Table \ref{tab:model_metrics} and Figure \ref{fig:cv_dashboards}). At $N=1000$ the average correlation of the out-of-sample CATEs is $r = 0.69$; at $N=4000$ it is $r = 0.81$, and $r = 0.84$ when both models are trained on the whole CANVAS-R sample. Spearman correlations are similar. This indicates similar estimated treatment effect patterns, although agreement between models does not establish the true functional form. At smaller sample sizes, both models often shrink the treatment effect back towards homogeneity, BCF more strongly.

The uplift curves in Figure \ref{fig:cv_dashboards} are comparable for both models and remain close to the diagonal representing random targeting. This limits the evidence for improved targeting despite agreement between the models. The normalized net areas under the uplift curve in Table \ref{tab:model_metrics} are similar as well. Thus, even when the CATE magnitudes differ at lower sample sizes, the models rank patients similarly. B-SLIM also has smaller RMSEs for the out-of-sample outcome ($Y$) at every training size in Table \ref{tab:model_metrics}. In this evaluation, B-SLIM therefore provides comparable ranking while also allowing direct variable selection, which we examine in the next section.

To standardize the area under the uplift curve, write $U(f)$ for cumulative uplift when the fraction $f$ of the validation sample is selected. With $n_v=3004$, we report
\[
 \frac{\operatorname{NetAUUC}}{n_v}
 =\frac{1}{n_v}\left\{\int_0^1 U(f)\,df-\frac{U(1)}{2}\right\}.
\]
We subtract the area under the straight line for random targeting and divide by the validation-sample size. Positive values indicate greater cumulative HbA1c reduction than random targeting. Table \ref{tab:model_metrics} reports this normalized measure.

At $N=500$, the outcome-RMSE standard deviation is much lower for B-SLIM than for BCF ($0.016$ versus $0.119$). We interpret this difference as reflecting the greater sample-to-sample variation induced by BCF's tree-based treatment-effect structure.

\begin{figure}[htbp]
    \centering
    \input{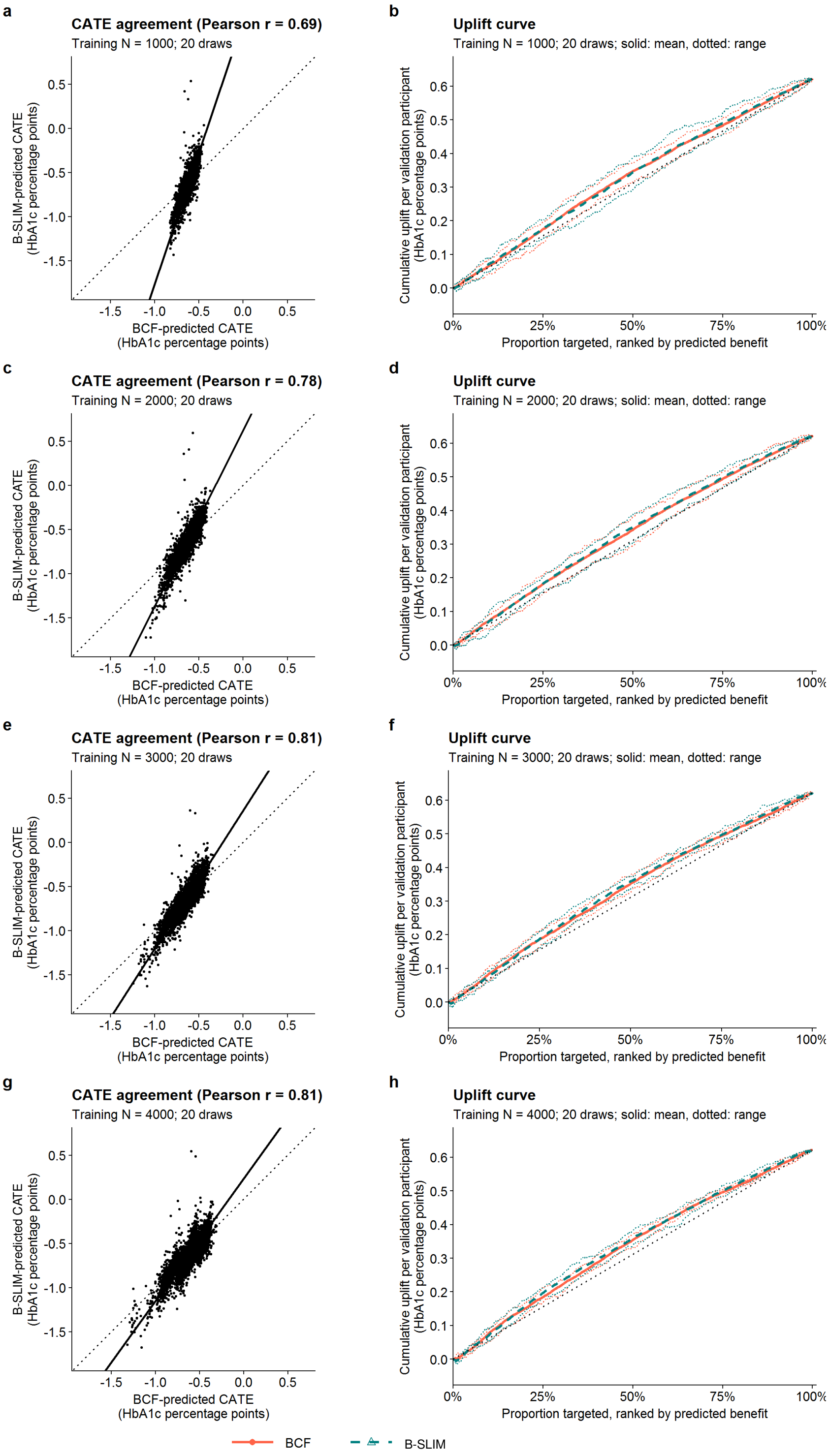}
    \caption{Out-of-sample CATE agreement (left) and uplift curves (right) for B-SLIM and BCF at $N=1000$, $2000$, $3000$, and $4000$ (mean over 20 draws). The orthogonal fit is a total least-squares line between the two models' predictions, minimizing perpendicular distances; it is unrelated to treatment residualisation in Section \ref{subsec:ortho_supp}. Dotted min--max lines show the pointwise range across the 20 training draws. The dotted diagonal marks equality in the agreement panels and random targeting in the uplift panels.}
    \label{fig:cv_dashboards}
\end{figure}

\begin{table}[htbp]
\centering
\caption{Out-of-sample performance on CANVAS. Values are mean $\pm$ standard deviation over 20 training draws, except the last row, which is a single fit on the full CANVAS-R sample. RMSE is for the outcome $Y$. Correlation and Spearman are between the two models' out-of-sample CATEs. Net AUUC subtracts the random-targeting area and is divided by the 3004 validation participants; its units are HbA1c change.}
\label{tab:model_metrics}
\resizebox{\textwidth}{!}{%
\begin{tabular}{lccccccc}
\toprule
& & \multicolumn{2}{c}{Outcome RMSE} & \multicolumn{2}{c}{CATE agreement} & \multicolumn{2}{c}{Net AUUC/$n_v$}\\
\cmidrule(lr){3-4}\cmidrule(lr){5-6}\cmidrule(lr){7-8}
$N$ & Draws & BCF & B-SLIM & $r$ & Spearman & BCF & B-SLIM\\
\midrule
500 & 20 & $1.009 \pm 0.119$ & $0.947 \pm 0.016$ & $0.702 \pm 0.137$ & $0.759 \pm 0.088$ & $0.009 \pm 0.013$ & $0.012 \pm 0.015$ \\
1000 & 20 & $0.974 \pm 0.078$ & $0.926 \pm 0.009$ & $0.692 \pm 0.130$ & $0.727 \pm 0.108$ & $0.019 \pm 0.010$ & $0.021 \pm 0.014$ \\
1500 & 20 & $0.952 \pm 0.057$ & $0.918 \pm 0.004$ & $0.741 \pm 0.102$ & $0.778 \pm 0.090$ & $0.020 \pm 0.009$ & $0.025 \pm 0.008$ \\
2000 & 20 & $0.979 \pm 0.044$ & $0.916 \pm 0.005$ & $0.778 \pm 0.088$ & $0.803 \pm 0.074$ & $0.023 \pm 0.009$ & $0.026 \pm 0.010$ \\
3000 & 20 & $0.970 \pm 0.058$ & $0.913 \pm 0.002$ & $0.811 \pm 0.069$ & $0.828 \pm 0.052$ & $0.025 \pm 0.004$ & $0.028 \pm 0.006$ \\
4000 & 20 & $0.944 \pm 0.024$ & $0.910 \pm 0.002$ & $0.813 \pm 0.033$ & $0.823 \pm 0.031$ & $0.027 \pm 0.003$ & $0.031 \pm 0.005$ \\
5018 & 1 & $0.950$ & $0.909$ & $0.843$ & $0.851$ & $0.029$ & $0.035$ \\
\bottomrule
\end{tabular}}
\end{table}
\newpage

\subsection{Interpretability and Variable Selection in Practice}
\label{subsec:applied_interpretability}
The out-of-sample evaluations show comparable ranking for the two models. B-SLIM also allows structural interpretation. We therefore applied the Shapley value decomposition to the CANVAS-R data to describe the covariate contributions to treatment effect heterogeneity. The Shapley summaries and the HbA1c and age coefficient posteriors are shown in the main text (Figures \ref{fig:shapley_contribution} and \ref{fig:posterior_HBA1C}). Table \ref{tab:canvas_covariates} defines the abbreviations.

\begin{table}[htbp]
\centering\small
\caption{Baseline covariates in the CANVAS analysis and figure labels.}
\label{tab:canvas_covariates}
\begin{tabular}{lp{0.73\textwidth}}
\toprule
Label & Covariate\\
\midrule
BLBMI & Body mass index\\
AGE & Age\\
HHFFL & History of heart failure\\
BLSBP & Systolic blood pressure\\
SEX & Sex\\
DIADUR & Diabetes duration\\
HCEREBFL & History of cerebrovascular disease\\
BLA1C & HbA1c\\
BLGFR & Estimated glomerular filtration rate\\
INSULBL & Insulin use\\
GLP1BL & GLP-1 receptor agonist use\\
BLACR & Urinary albumin-to-creatinine ratio\\
RACE & Recorded race, represented by sum contrasts RACE1--RACE5\\
METFORBL & Metformin use\\
RANGIBL & Use of agents acting on the renin--angiotensin system\\
\bottomrule
\end{tabular}
\end{table}

For RACE, the signed column contributions are summed within each individual before taking absolute values. RACE, baseline HbA1c, and age have the largest contributions, predominantly through main effects. The recorded race categories are White; Black or African American; Asian; American Indian or Alaska Native; Native Hawaiian or Other Pacific Islander; and Other/Multiple. We combine Other and Multiple, and treat Unknown and Not Reported as missing before complete-case filtering.

To assess each covariate, we compare the full fitted CATE, including its interactions, at selected values while averaging over the empirical distribution of the other covariates. For continuous covariates, we compare the 90th and 10th percentiles; for binary covariates, the two levels; and for RACE, every pair of categories. For posterior contrast draws $d_\ell^{(m)}$, with posterior mean $\bar d_\ell$ and standard deviation $s_\ell$, we take the $95$th percentile $k$ of $\max_\ell |(d_\ell^{(m)}-\bar d_\ell)/s_\ell|$. The intervals $\bar d_\ell\pm k s_\ell$ are simultaneous across all assessed contrasts. A covariate is flagged when at least one of its intervals excludes zero.

Figure \ref{fig:canvas_full_contrasts} shows the simultaneous intervals for all assessed contrasts.

\begin{figure}[p]
    \centering
    \includegraphics[width=\textwidth,height=0.78\textheight,keepaspectratio]{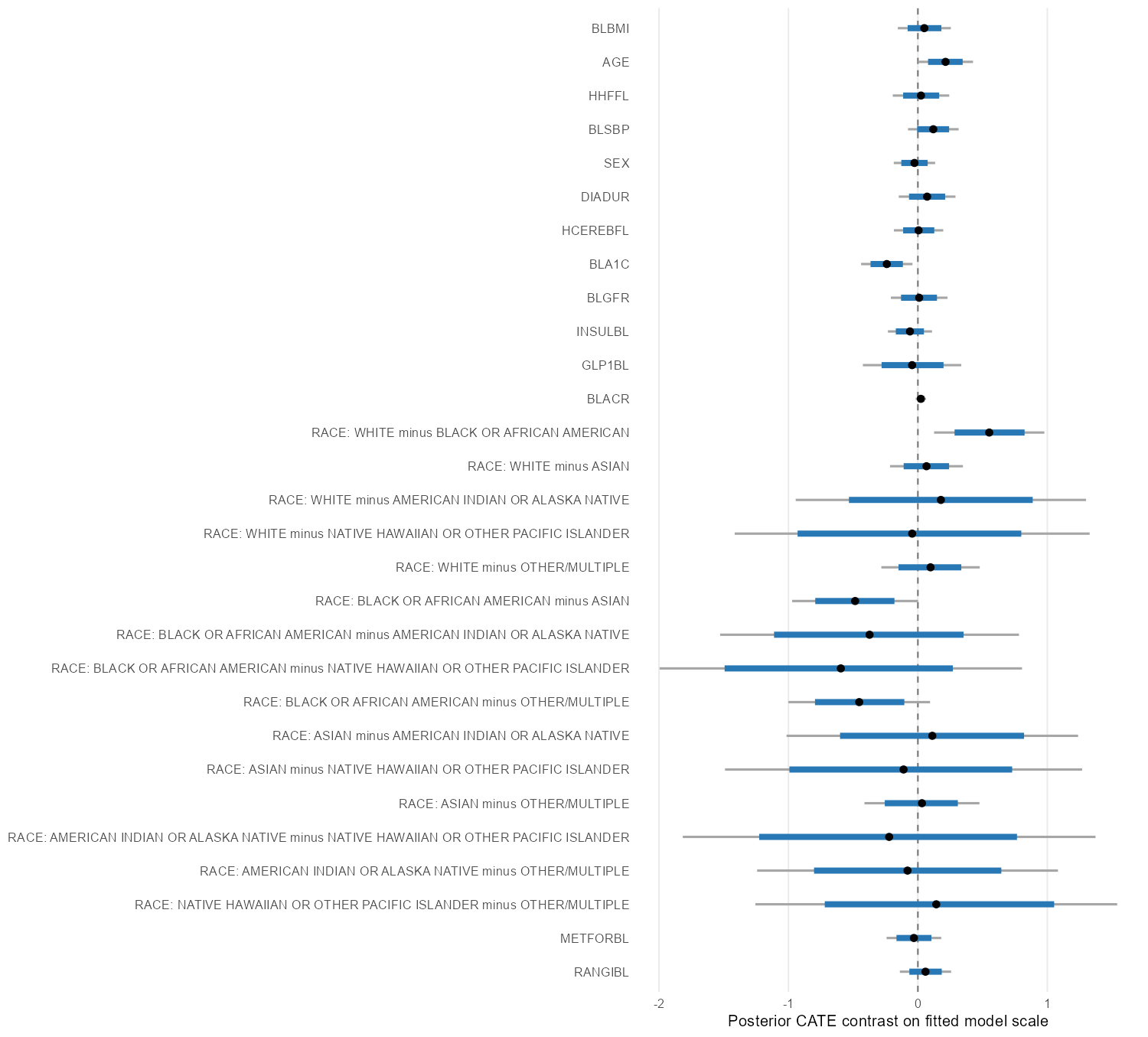}
    \caption{Full fitted CATE contrasts in CANVAS-R, including interactions and averaged over the other covariates. Continuous covariates compare the 90th minus 10th percentile; binary covariates compare the higher-coded minus lower-coded level; RACE compares the first named minus second named category. Points are posterior means; blue and grey intervals are pointwise and simultaneous $95\%$ credible intervals, respectively. Contrasts are on the fitted model scale.}
    \label{fig:canvas_full_contrasts}
\end{figure}
\FloatBarrier

\section{Ranking Performance in Simulation}
\label{subsec:ranking_supp}

Tables \ref{tab:med_ranking} and \ref{tab:mild_ranking} display interval-based selection performance under the Heavy Saturation and Mild Saturation scenarios, respectively.

Two patterns hold across both scenarios. The advantage of B-SLIM is concentrated at the sharp thresholds. Under heavy saturation it attains both higher sensitivity and higher precision than BCF at the top 10\% threshold, recovering on average 29.9\% of the true top decile against 1.3\% for BCF, at a mean precision of 0.994 against 0.757. At the top 50\% threshold under heavy saturation, mean sensitivities are similar (0.706 for BCF against 0.686), while their mean precisions are close (0.998 against 0.998). The advantage under this selection rule is therefore concentrated at the more selective threshold.

Table \ref{tab:med_ranking} shows that under heavy saturation the precision of B-SLIM stays at or above 0.956 at every sample size and threshold, while BCF's top 10\% precision ranges from 0.397 to 0.965 and is weakest when $\mu(x)$ is non-linear. The selection rule can therefore identify high-benefit individuals in this monotone simulation, even though the CATE magnitudes are biased and coverage drops (Table \ref{tab:med_sat_scenario}). Under mild saturation (Table \ref{tab:mild_ranking}) the same ordering at the top decile shows up from $N \ge 1000$ onwards, but at $N = 250$ neither model is dependable at the top decile: precision falls to 0.748 for B-SLIM under a linear $\mu(x)$ and to 0.261 for BCF under a non-linear one. At small sample sizes, precision at the sharpest threshold is not stable. These ranking results are for a monotone treatment function.

\begin{table}[ht]
\centering
\caption{Interval-based selection performance: heavy saturation scenario. A patient is flagged if the lower bound of the 95\% credible interval exceeds the threshold. Precision is averaged only over replicates in which the model flagged at least one patient, so it should be read together with sensitivity.}
\label{tab:med_ranking}
\resizebox{\textwidth}{!}{
\begin{tabular}{lr cc cc cc c cc cc cc}
\toprule
& & \multicolumn{6}{c}{\textbf{ BCF}} & & \multicolumn{6}{c}{\textbf{B-SLIM}} \\
\cmidrule{3-8} \cmidrule{10-15}
$\mu(x)$ Form & $N$ & \multicolumn{2}{c}{Top 10\%} & \multicolumn{2}{c}{Top 25\%} & \multicolumn{2}{c}{Top 50\%} & & \multicolumn{2}{c}{Top 10\%} & \multicolumn{2}{c}{Top 25\%} & \multicolumn{2}{c}{Top 50\%} \\
\cmidrule{3-4} \cmidrule{5-6} \cmidrule{7-8} \cmidrule{10-11} \cmidrule{12-13} \cmidrule{14-15}
& & Sens & Prec & Sens & Prec & Sens & Prec & & Sens & Prec & Sens & Prec & Sens & Prec \\
\midrule
\multirow{6}{*}{Linear}
& 250 & 0.016 & 0.920 & 0.041 & 0.983 & 0.427 & 0.997 &  & 0.101 & 0.989 & 0.102 & 0.999 & 0.406 & 0.997 \\
& 500 & 0.014 & 0.905 & 0.065 & 0.997 & 0.633 & 1.000 &  & 0.187 & 0.997 & 0.185 & 1.000 & 0.594 & 0.999 \\
& 750 & 0.018 & 0.934 & 0.115 & 0.995 & 0.708 & 1.000 &  & 0.251 & 0.998 & 0.232 & 1.000 & 0.666 & 0.999 \\
& 1000 & 0.025 & 0.916 & 0.150 & 0.995 & 0.747 & 0.999 &  & 0.289 & 1.000 & 0.268 & 1.000 & 0.722 & 0.999 \\
& 1500 & 0.014 & 0.827 & 0.227 & 0.996 & 0.795 & 0.999 &  & 0.349 & 1.000 & 0.314 & 1.000 & 0.771 & 0.999 \\
& 3000 & 0.013 & 0.965 & 0.468 & 0.994 & 0.848 & 0.999 &  & 0.460 & 1.000 & 0.388 & 1.000 & 0.841 & 0.999 \\
\midrule
\multirow{6}{*}{Non-linear}
& 250 & 0.013 & 0.397 & 0.032 & 0.844 & 0.471 & 0.992 &  & 0.146 & 0.956 & 0.140 & 0.986 & 0.483 & 0.989 \\
& 500 & 0.014 & 0.553 & 0.079 & 0.919 & 0.685 & 0.997 &  & 0.240 & 0.989 & 0.231 & 0.999 & 0.648 & 0.997 \\
& 750 & 0.006 & 0.602 & 0.104 & 0.972 & 0.726 & 0.998 &  & 0.298 & 1.000 & 0.271 & 1.000 & 0.705 & 0.999 \\
& 1000 & 0.006 & 0.486 & 0.170 & 0.975 & 0.759 & 0.999 &  & 0.342 & 0.999 & 0.304 & 1.000 & 0.745 & 0.999 \\
& 1500 & 0.006 & 0.785 & 0.281 & 0.995 & 0.814 & 0.999 &  & 0.415 & 1.000 & 0.355 & 1.000 & 0.795 & 0.999 \\
& 3000 & 0.013 & 0.793 & 0.507 & 0.997 & 0.860 & 0.998 &  & 0.504 & 1.000 & 0.414 & 1.000 & 0.855 & 1.000 \\
\bottomrule
\end{tabular}}
\end{table}

\begin{table}[ht]
\centering
\caption{Interval-based selection performance: mild saturation scenario. A patient is flagged if the lower bound of the 95\% credible interval exceeds the threshold. Precision is averaged only over replicates in which the model flagged at least one patient, so it should be read together with sensitivity.}
\label{tab:mild_ranking}
\resizebox{\textwidth}{!}{
\begin{tabular}{lr cc cc cc c cc cc cc}
\toprule
& & \multicolumn{6}{c}{\textbf{BCF}} & & \multicolumn{6}{c}{\textbf{B-SLIM}} \\
\cmidrule{3-8} \cmidrule{10-15}
$\mu(x)$ Form & $N$ & \multicolumn{2}{c}{Top 10\%} & \multicolumn{2}{c}{Top 25\%} & \multicolumn{2}{c}{Top 50\%} & & \multicolumn{2}{c}{Top 10\%} & \multicolumn{2}{c}{Top 25\%} & \multicolumn{2}{c}{Top 50\%} \\
\cmidrule{3-4} \cmidrule{5-6} \cmidrule{7-8} \cmidrule{10-11} \cmidrule{12-13} \cmidrule{14-15}
& & Sens & Prec & Sens & Prec & Sens & Prec & & Sens & Prec & Sens & Prec & Sens & Prec \\
\midrule
\multirow{6}{*}{Linear}
& 250 & 0.009 & 0.649 & 0.011 & 0.876 & 0.023 & 0.953 &  & 0.026 & 0.748 & 0.032 & 0.863 & 0.059 & 0.913 \\
& 500 & 0.010 & 0.958 & 0.015 & 0.961 & 0.048 & 0.990 &  & 0.039 & 0.885 & 0.059 & 0.969 & 0.115 & 0.991 \\
& 750 & 0.014 & 0.902 & 0.032 & 0.955 & 0.080 & 0.992 &  & 0.065 & 0.945 & 0.090 & 0.981 & 0.170 & 0.992 \\
& 1000 & 0.026 & 0.901 & 0.045 & 0.956 & 0.116 & 0.992 &  & 0.058 & 0.974 & 0.105 & 0.988 & 0.224 & 0.992 \\
& 1500 & 0.017 & 0.893 & 0.060 & 0.978 & 0.190 & 0.996 &  & 0.080 & 0.989 & 0.156 & 0.992 & 0.314 & 0.996 \\
& 3000 & 0.035 & 0.974 & 0.150 & 0.991 & 0.394 & 0.998 &  & 0.182 & 0.991 & 0.298 & 0.996 & 0.497 & 0.997 \\
\midrule
\multirow{6}{*}{Non-linear}
& 250 & 0.011 & 0.261 & 0.014 & 0.468 & 0.034 & 0.742 &  & 0.040 & 0.484 & 0.053 & 0.706 & 0.088 & 0.885 \\
& 500 & 0.011 & 0.232 & 0.019 & 0.563 & 0.060 & 0.860 &  & 0.053 & 0.854 & 0.085 & 0.926 & 0.157 & 0.970 \\
& 750 & 0.004 & 0.333 & 0.017 & 0.761 & 0.075 & 0.940 &  & 0.059 & 0.959 & 0.101 & 0.987 & 0.206 & 0.991 \\
& 1000 & 0.006 & 0.468 & 0.028 & 0.822 & 0.127 & 0.959 &  & 0.073 & 0.971 & 0.136 & 0.987 & 0.271 & 0.994 \\
& 1500 & 0.011 & 0.629 & 0.059 & 0.895 & 0.215 & 0.989 &  & 0.123 & 0.987 & 0.212 & 0.992 & 0.379 & 0.996 \\
& 3000 & 0.030 & 0.810 & 0.144 & 0.979 & 0.429 & 0.997 &  & 0.226 & 0.995 & 0.354 & 0.997 & 0.544 & 0.999 \\
\bottomrule
\end{tabular}}
\end{table}

\FloatBarrier
\section{Computational Details}
\label{sec:full_conditionals}
\label{sec:computational_details}

We use Bayesian backfitting to alternate between the prognostic and treatment
functions. Let $u_i$ denote the fixed treatment coding, with $u_i=z_i$ for
$0/1$ coding and $u_i=z_i-1/2$ for symmetric coding. Let $h_i$ contain the
$p$ covariate columns and the $q$ included interaction columns, and collect
their coefficients in $b$. The treatment-weighted design $W$ has rows
$u_i h_i^\top$, so that $y=\mu+u\alpha+Wb+\varepsilon$ and
$\tau(x_i)=\alpha+h_i^\top b$. Write $\mathcal P$ for the included interaction pairs, $K=p+q$, and
\[
 D=\operatorname{diag}\left(\lambda_1^2,\ldots,\lambda_p^2,
       \{\lambda_{j,k}^2\lambda_{\mathrm{glob}}^2\}_{(j,k)\in\mathcal P}\right),
 \qquad Q=W^\top W+D^{-1}.
\]
All quantities below are on the fitted scale, after centering the outcome and dividing it by its training-sample standard deviation. The simulations use the generated covariates, while IHDP and CANVAS standardize continuous covariates before forming interaction products. Binary and categorical contrast columns and the treatment coding are not standardized. We use the inverse-gamma
parameterization $p(v)\propto v^{-a-1}\exp(-b/v)$ for
$v\sim\operatorname{IG}(a,b)$; conditioning on the data and all other
current parameters is denoted by $\mid -$.

For the prognostic function, we use the BART prior of
Chipman et~al.,\citep{chipman2010bart} with $T=250$ trees and splitting probability
$0.95(1+d)^{-2}$ at depth $d$, a maximum depth of 10, and at least five
observations per leaf. Given leaf variance $s_\mu^2$, the leaf parameters
have independent $\mathcal N(0,s_\mu^2)$ priors. For a leaf containing
$n_{t\ell}$ observations, let $S_{t\ell}$ be the sum of the tree-specific
partial residuals defined in Algorithm~\ref{algorithm backfitting}. Then
\begin{equation}
 v_{t\ell}=\left(\frac{n_{t\ell}}{\sigma^2}+\frac{1}{s_\mu^2}\right)^{-1},
 \qquad
 \theta_{t\ell}\mid -\sim
 \mathcal N\left(\frac{v_{t\ell}S_{t\ell}}{\sigma^2},v_{t\ell}\right).
 \label{eq:fc_leaf}
\end{equation}
With $s_\mu^2\sim\operatorname{IG}(a_\mu/2,b_\mu/2)$ and $L$ leaves
across the forest, the leaf variance is drawn from
\begin{equation}
 s_\mu^2\mid -\sim\operatorname{IG}\left(
 \frac{a_\mu+L}{2},\frac{b_\mu+\sum_{t,\ell}\theta_{t\ell}^2}{2}\right),
 \qquad a_\mu=3,\quad b_\mu=1/T.
 \label{eq:fc_leaf_scale}
\end{equation}

The treatment intercept and coefficients have the Gaussian full conditionals
\begin{align}
 v_\alpha&=\left(\frac{u^\top u}{\sigma^2}+\frac{1}{100}\right)^{-1},
 &\alpha\mid -&\sim\mathcal N\left(
 \frac{v_\alpha u^\top(y-\mu-Wb)}{\sigma^2},v_\alpha\right),
 \label{eq:fc_alpha}\\
 b\mid -&\sim\mathcal N_K\left(
 Q^{-1}W^\top(y-\mu-u\alpha),\;\sigma^2Q^{-1}\right).
 \label{eq:fc_beta}
\end{align}
Following Makalic and Schmidt,\citep{makalic2015simple} a unit half-Cauchy scale is represented
by $\lambda^2\mid\nu\sim\operatorname{IG}(1/2,1/\nu)$ and
$\nu\sim\operatorname{IG}(1/2,1)$. Applying this to each local scale and
using auxiliary variable $\xi$ for the global scale gives
\begin{align}
 \nu_j\mid -&\sim\operatorname{IG}(1,1+\lambda_j^{-2}),
 &\lambda_j^2\mid -&\sim\operatorname{IG}\left(
 1,\frac{1}{\nu_j}+\frac{\beta_j^2}{2\sigma^2}\right),
 \label{eq:fc_lambda_main}\\
 \nu_{j,k}\mid -&\sim\operatorname{IG}(1,1+\lambda_{j,k}^{-2}),
 &\lambda_{j,k}^2\mid -&\sim\operatorname{IG}\left(
 1,\frac{1}{\nu_{j,k}}+
 \frac{\beta_{j,k}^2}{2\sigma^2\lambda_{\mathrm{glob}}^2}\right),
 \label{eq:fc_lambda_int}\\
 \xi\mid -&\sim\operatorname{IG}(1,1+\lambda_{\mathrm{glob}}^{-2}),
 &\lambda_{\mathrm{glob}}^2\mid -&\sim\operatorname{IG}\left(
 \frac{q+1}{2},\frac{1}{\xi}+\frac{1}{2\sigma^2}
 \sum_{(j,k)\in\mathcal P}\frac{\beta_{j,k}^2}{\lambda_{j,k}^2}\right).
 \label{eq:fc_global}
\end{align}
The main effects have local scales only; the global scale applies to the
interaction block. Finally, with total residuals $e=y-\mu-u\alpha-Wb$,
\begin{equation}
 \sigma^2\mid -\sim\operatorname{IG}\left(
 1+\frac{N+K}{2},\;0.001+\frac{e^\top e+b^\top D^{-1}b}{2}\right).
 \label{eq:fc_sigma}
\end{equation}
The intercept and leaf priors are independent of $\sigma^2$; the
regularized coefficient priors contribute the additional $K/2$ and
$b^\top D^{-1}b/2$ terms.

\begin{algorithm}[htbp]
\caption{Bayesian Backfitting MCMC Scheme for B-SLIM}
\label{algorithm backfitting}
\begin{algorithmic}[1]
\State \textbf{Initialize:} Prognostic forest $\mu(\cdot)$, leaf variance
$s_\mu^2$, linear parameters $\alpha,b$, local and global shrinkage
parameters, auxiliary variables, and noise variance $\sigma^2$.
\State Fix the covariates, interactions, treatment coding $u$, and propensity
score (fitted in observational data or known from the randomized design).
\For{iteration $m=1,\dots,N_{\mathrm{burn}}+N_{\mathrm{keep}}$}
    \State \textbf{1. Update Prognostic Function ($\mu$):}
    \State Compute partial residuals for $i=1,\ldots,N$:
    \State \quad $R_i^*=y_i-u_i\tau^{(m-1)}(x_i)$.
    \For{each tree $t=1,\ldots,T$}
        \State Compute tree-specific residuals
        $R^*_{i,(-t)}=R_i^*-\sum_{s\ne t}g_s(x_i,\hat\pi_i)$.
        \State Propose a tree modification (e.g., grow, prune) and
        accept/reject via Metropolis--Hastings, integrating out leaf values.
        \State Draw the leaf parameters conditional on the updated tree
        structure and $R^*_{i,(-t)}$ using \eqref{eq:fc_leaf}.
    \EndFor
    \State Draw the common leaf variance using \eqref{eq:fc_leaf_scale}.
    \State \textbf{2. Update Treatment Function ($\tau$):}
    \State Compute partial residuals for $i=1,\ldots,N$:
    \State \quad $R_i^{**}=y_i-\mu^{(m)}(x_i,\hat\pi_i)$.
    \State Draw the intercept $\alpha$ using \eqref{eq:fc_alpha}.
    \State Draw the main-effect and interaction coefficients jointly in $b$
    using \eqref{eq:fc_beta}, with residuals $R^{**}-u\alpha$ and design $W$.
    \State Draw each local auxiliary and then its local scale using
    \eqref{eq:fc_lambda_main} and \eqref{eq:fc_lambda_int}.
    \State Draw the global auxiliary $\xi$ and then
    $\lambda_{\mathrm{glob}}^2$ using \eqref{eq:fc_global}.
    \State Update the CATE function:
    \State \quad $\tau^{(m)}(x_i)=\alpha+\sum_{j=1}^{p}\beta_j x_{i,j}+\sum_{(j,k)\in\mathcal P}\beta_{j,k}x_{i,j}x_{i,k}$.
    \State \textbf{3. Update Noise Variance ($\sigma^2$):}
    \State Compute total residuals:
    \State \quad $e_i=y_i-\mu^{(m)}(x_i,\hat\pi_i)-u_i\tau^{(m)}(x_i)$.
    \State Draw $\sigma^{2(m)}$ using \eqref{eq:fc_sigma}, including the
    coefficient-prior term $b^\top D^{-1}b$.
    \If{$m>N_{\mathrm{burn}}$}
        \State Store the draw and the predictions of interest.
    \EndIf
\EndFor
\end{algorithmic}
\end{algorithm}

The latest available parameter values are used at each step.

\FloatBarrier

\clearpage
\endgroup

\bibliographystyle{sim-ama}
\bibliography{references-8.3.8}

\begin{thebibliography}{10}
\providecommand \doibase [0]{https://doi.org/}%

\bibitem{sparano2018adjuvant}
Sparano JA, Gray RJ, Makower DF, et al. Adjuvant chemotherapy guided by a
  21-gene expression assay in breast cancer. {\it N Engl J Med.}
  2018\string;379(2)\string:111--121.
\newblock \href {\doibase 10.1056/nejmoa1804710} {doi: 10.1056/nejmoa1804710}

\bibitem{rubin1974estimating}
Rubin DB. Estimating causal effects of treatments in randomized and
  nonrandomized studies. {\it J Educ Psychol.}
  1974\string;66(5)\string:688--701.
\newblock \href {\doibase 10.1037/h0037350} {doi: 10.1037/h0037350}

\bibitem{kent2020predictive}
Kent DM, Paulus JK, van Klaveren D, et al. The predictive approaches to
  treatment effect heterogeneity ({PATH}) statement. {\it Ann Intern Med.}
  2020\string;172(1)\string:35--45.
\newblock \href {\doibase 10.7326/m18-3667} {doi: 10.7326/m18-3667}

\bibitem{tian2014simple}
Tian L, Alizadeh AA, Gentles AJ, Tibshirani R. A simple method for estimating
  interactions between a treatment and a large number of covariates. {\it J Am
  Stat Assoc.} 2014\string;109(508)\string:1517--1532.
\newblock \href {\doibase 10.1080/01621459.2014.951443} {doi:
  10.1080/01621459.2014.951443}

\bibitem{curth2021inductive}
Curth A, van~der Schaar M. On inductive biases for heterogeneous treatment
  effect estimation. In: {\it Advances in Neural Information Processing
  Systems}. Vol. 34. Curran Associates, Inc.; 2021\string:15883--15894.
\newblock
  \url{https://proceedings.neurips.cc/paper_files/paper/2021/hash/8526e0962a844e4a2f158d831d5fddf7-Abstract.html}.

\bibitem{athey2016recursive}
Athey S, Imbens G. Recursive partitioning for heterogeneous causal effects.
  {\it Proc Natl Acad Sci U S A.} 2016\string;113(27)\string:7353--7360.
\newblock \href {\doibase 10.1073/pnas.1510489113} {doi:
  10.1073/pnas.1510489113}

\bibitem{wager2018estimation}
Wager S, Athey S. Estimation and inference of heterogeneous treatment effects
  using random forests. {\it J Am Stat Assoc.}
  2018\string;113(523)\string:1228--1242.
\newblock \href {\doibase 10.1080/01621459.2017.1319839} {doi:
  10.1080/01621459.2017.1319839}

\bibitem{breiman2001random}
Breiman L. Random forests. {\it Mach Learn.} 2001\string;45(1)\string:5--32.
\newblock \href {\doibase 10.1023/a:1010933404324} {doi:
  10.1023/a:1010933404324}

\bibitem{hahn2020bayesian}
Hahn PR, Murray JS, Carvalho CM. {Bayesian} regression tree models for causal
  inference: Regularization, confounding, and heterogeneous effects (with
  discussion). {\it Bayesian Anal.} 2020\string;15(3)\string:965--1056.
\newblock \href {\doibase 10.1214/19-ba1195} {doi: 10.1214/19-ba1195}

\bibitem{chipman2010bart}
Chipman HA, George EI, McCulloch RE. {BART}: {Bayesian} additive regression
  trees. {\it Ann Appl Stat.} 2010\string;4(1)\string:266--298.
\newblock \href {\doibase 10.1214/09-aoas285} {doi: 10.1214/09-aoas285}

\bibitem{dorie2019automated}
Dorie V, Hill J, Shalit U, Scott M, Cervone D. Automated versus do-it-yourself
  methods for causal inference: Lessons learned from a data analysis
  competition. {\it Stat Sci.} 2019\string;34(1)\string:43--68.
\newblock \href {\doibase 10.1214/18-sts667} {doi: 10.1214/18-sts667}

\bibitem{svensson2026overview}
Svensson D, Hermansson E, Nikolaou N, Sechidis K, Lipkovich I. Overview and
  practical recommendations on using {Shapley} values for identifying
  predictive biomarkers via {CATE} modeling. {\it Stat Med.}
  2026\string;45(1--2)\string:e70375.
\newblock \href {\doibase 10.1002/sim.70375} {doi: 10.1002/sim.70375}

\bibitem{zeldow2019semiparametric}
Zeldow B, Lo~Re V III, Roy J. A semiparametric modeling approach using
  {Bayesian} additive regression trees with an application to evaluate
  heterogeneous treatment effects. {\it Ann Appl Stat.}
  2019\string;13(3)\string:1989--2010.
\newblock \href {\doibase 10.1214/19-aoas1266} {doi: 10.1214/19-aoas1266}

\bibitem{prado2025shared}
Prado EB, Parnell AC, Moral RA, McJames N, O'Shea A, Murphy K. Accounting for
  shared covariates in semiparametric {Bayesian} additive regression trees.
  {\it Ann Appl Stat.} 2025\string;19(1)\string:302--328.
\newblock \href {\doibase 10.1214/24-aoas1960} {doi: 10.1214/24-aoas1960}

\bibitem{carvalho2010horseshoe}
Carvalho CM, Polson NG, Scott JG. The horseshoe estimator for sparse signals.
  {\it Biometrika.} 2010\string;97(2)\string:465--480.
\newblock \href {\doibase 10.1093/biomet/asq017} {doi: 10.1093/biomet/asq017}

\bibitem{van2014horseshoe}
van~der Pas SL, Kleijn BJK, van~der Vaart AW. The horseshoe estimator:
  Posterior concentration around nearly black vectors. {\it Electron J Stat.}
  2014\string;8(2)\string:2585--2618.
\newblock \href {\doibase 10.1214/14-ejs962} {doi: 10.1214/14-ejs962}

\bibitem{hastie2000bayesian}
Hastie T, Tibshirani R. {Bayesian} backfitting (with comments and a rejoinder
  by the authors). {\it Stat Sci.} 2000\string;15(3)\string:196--223.
\newblock \href {\doibase 10.1214/ss/1009212815} {doi: 10.1214/ss/1009212815}

\bibitem{imbens2015causal}
Imbens GW, Rubin DB. {\it Causal Inference for Statistics, Social, and
  Biomedical Sciences: An Introduction}.
\newblock Cambridge University Press; 2015.
\newblock \url{https://doi.org/10.1017/cbo9781139025751}

\bibitem{linero2024nonparametric}
Linero AR. In nonparametric and high-dimensional models, {Bayesian}
  ignorability is an informative prior. {\it J Am Stat Assoc.}
  2024\string;119(548)\string:2785--2798.
\newblock \href {\doibase 10.1080/01621459.2023.2278202} {doi:
  10.1080/01621459.2023.2278202}

\bibitem{robinson1988root}
Robinson PM. Root-{N}-consistent semiparametric regression. {\it Econometrica.}
  1988\string;56(4)\string:931--954.
\newblock \href {\doibase 10.2307/1912705} {doi: 10.2307/1912705}

\bibitem{makalic2015simple}
Makalic E, Schmidt DF. A simple sampler for the horseshoe estimator. {\it IEEE
  Signal Process Lett.} 2016\string;23(1)\string:179--182.
\newblock \href {\doibase 10.1109/lsp.2015.2503725} {doi:
  10.1109/lsp.2015.2503725}

\bibitem{van2024linked}
van~de Wiel MA, Amestoy M, Hoogland J. Linked shrinkage to improve estimation
  of interaction effects in regression models. {\it Epidemiol Methods.}
  2024\string;13(1)\string:20230039.
\newblock \href {\doibase 10.1515/em-2023-0039} {doi: 10.1515/em-2023-0039}

\bibitem{ihdp1990trial}
{Infant Health and Development Program}. Enhancing the outcomes of
  low-birth-weight, premature infants: A multisite, randomized trial. {\it
  JAMA.} 1990\string;263(22)\string:3035--3042.
\newblock \href {\doibase 10.1001/jama.1990.03440220059030} {doi:
  10.1001/jama.1990.03440220059030}

\bibitem{hill2011bayesian}
Hill JL. {Bayesian} nonparametric modeling for causal inference. {\it J Comput
  Graph Stat.} 2011\string;20(1)\string:217--240.
\newblock \href {\doibase 10.1198/jcgs.2010.08162} {doi:
  10.1198/jcgs.2010.08162}

\bibitem{ihdpbenchmarkdata}
Johansson FD. {IHDP} semi-synthetic benchmark data: 100 replications. Public
  train and test data files.
\newblock Accessed November 5, 2026.
  \url{https://www.fredjo.com/files/ihdp_npci_1-100.train.npz};
  \url{https://www.fredjo.com/files/ihdp_npci_1-100.test.npz}.

\bibitem{curth2021really}
Curth A, Svensson D, Weatherall J, van~der Schaar M. Really Doing Great at
  Estimating {CATE}? A Critical Look at {ML} Benchmarking Practices in
  Treatment Effect Estimation. In: {\it Proceedings of the Neural Information
  Processing Systems Track on Datasets and Benchmarks}. Vol. 1. 2021.
\newblock
  \url{https://datasets-benchmarks-proceedings.neurips.cc/paper/2021/hash/2a79ea27c279e471f4d180b08d62b00a-Abstract-round2.html}.

\bibitem{vanderlaan2007super}
van~der Laan MJ, Polley EC, Hubbard AE. Super learner. {\it Stat Appl Genet Mol
  Biol.} 2007\string;6(1).
\newblock \href {\doibase 10.2202/1544-6115.1309} {doi: 10.2202/1544-6115.1309}

\bibitem{kennedy2020optimal}
Kennedy EH. Towards optimal doubly robust estimation of heterogeneous causal
  effects. {\it Electron J Stat.} 2023\string;17(2)\string:3008--3049.
\newblock \href {\doibase 10.1214/23-ejs2157} {doi: 10.1214/23-ejs2157}

\bibitem{nie2021quasi}
Nie X, Wager S. Quasi-oracle estimation of heterogeneous treatment effects.
  {\it Biometrika.} 2021\string;108(2)\string:299--319.
\newblock \href {\doibase 10.1093/biomet/asaa076} {doi: 10.1093/biomet/asaa076}

\bibitem{kunzel2019metalearners}
K{\"u}nzel SR, Sekhon JS, Bickel PJ, Yu B. Metalearners for estimating
  heterogeneous treatment effects using machine learning. {\it Proc Natl Acad
  Sci U S A.} 2019\string;116(10)\string:4156--4165.
\newblock \href {\doibase 10.1073/pnas.1804597116} {doi:
  10.1073/pnas.1804597116}

\bibitem{neal2013rationale}
Neal B, Perkovic V, de~Zeeuw D, et al. Rationale, design, and baseline
  characteristics of the Canagliflozin Cardiovascular Assessment Study
  ({CANVAS}) - a randomized placebo-controlled trial. {\it Am Heart J.}
  2013\string;166(2)\string:217--223.e11.
\newblock \href {\doibase 10.1016/j.ahj.2013.05.007} {doi:
  10.1016/j.ahj.2013.05.007}

\bibitem{neal2017rationale}
Neal B, Perkovic V, Matthews DR, et al. Rationale, design and baseline
  characteristics of the CANagliflozin cardioVascular Assessment Study-Renal
  ({CANVAS-R}): A randomized, placebo-controlled trial. {\it Diabetes Obes
  Metab.} 2017\string;19(3)\string:387--393.
\newblock \href {\doibase 10.1111/dom.12829} {doi: 10.1111/dom.12829}

\bibitem{yodacanvasdata}
{Yale University Open Data Access Project}. {CANVAS} and {CANVAS-R} individual
  participant data. Controlled-access datasets, NCT01032629 and NCT01989754.
\newblock Access request 2025-0684. Accessed November 5, 2026.
  \url{https://yoda.yale.edu/data-request/2025-0684/}.

\bibitem{cefalu2013efficacy}
Cefalu WT, Leiter LA, Yoon KH, et al. Efficacy and safety of canagliflozin
  versus glimepiride in patients with type 2 diabetes inadequately controlled
  with metformin (CANTATA-SU): 52 week results from a randomised, double-blind,
  phase 3 non-inferiority trial. {\it Lancet.}
  2013\string;382(9896)\string:941--950.
\newblock \href {\doibase 10.1016/s0140-6736(13)60683-2} {doi:
  10.1016/s0140-6736(13)60683-2}

\bibitem{gavin2015racial}
Gavin JR III, Davies MJ, Davies M, Vijapurkar U, Alba M, Meininger G. The
  efficacy and safety of canagliflozin across racial groups in patients with
  type 2 diabetes mellitus. {\it Curr Med Res Opin.}
  2015\string;31(9)\string:1693--1702.
\newblock \href {\doibase 10.1185/03007995.2015.1067192} {doi:
  10.1185/03007995.2015.1067192}

\bibitem{chen2024ethnic}
Chen X, Hou X, Gao J, et al. Ethnic disparities in cardiovascular and renal
  responses to canagliflozin between {Asian} and {White} patients with type 2
  diabetes mellitus: A post hoc analysis of the {CANVAS} Program. {\it Diabetes
  Obes Metab.} 2024\string;26(3)\string:878--890.
\newblock \href {\doibase 10.1111/dom.15380} {doi: 10.1111/dom.15380}

\bibitem{bslimpaper}
Balci Y. {B-SLIM}: manuscript and analysis code. Research compendium; 2026.
\newblock Repository snapshot accessed November 5, 2026.
  \url{https://github.com/HYHBalci/bslim-paper}.

\bibitem{bslimsoftware}
Balci Y. {B-SLIM}: {BART} sparse linear interaction model for causal inference.
  R package source repository; 2026.
\newblock Repository snapshot accessed November 5, 2026.
  \url{https://github.com/HYHBalci/bslim}.

\bibitem{muller2025data}
M{\"u}ller MM, Bornkamp B, Bretz F, et al. Valid data-driven subgroup selection
  with uniform guarantees in clinical trials. {\it Stat Med.}
  2026\string;45(23--24)\string:e70745.
\newblock \href {\doibase 10.1002/sim.70745} {doi: 10.1002/sim.70745}

\bibitem{cheng2025chiseling}
Cheng N, Spector A, Janson L. Chiseling: Powerful and Valid Subgroup Selection
  via Interactive Machine Learning. arXiv preprint arXiv:2509.19490; 2025.
\newblock \url{https://doi.org/10.48550/arXiv.2509.19490}

\bibitem{chen2026comparing}
Chen Y, Sun S, Sechidis K, Zhang C, Hothorn T, Bornkamp B. Comparing Methods to
  Assess Treatment Effect Heterogeneity in General Parametric Regression
  Models. {\it Stat Med.} 2026\string;45(1--2)\string:e70381.
\newblock \href {\doibase 10.1002/sim.70381} {doi: 10.1002/sim.70381}

\bibitem{rzepakowski2010decision}
Rzepakowski P, Jaroszewicz S. Decision trees for uplift modeling. In: {\it 2010
  IEEE International Conference on Data Mining}. IEEE; 2010\string:441--450.
\newblock \url{https://doi.org/10.1109/icdm.2010.62}

\end{thebibliography}

\end{document}